\documentclass[12pt]{amsart}
\usepackage{amssymb,multicol}
\usepackage[mathscr]{eucal}

\usepackage{pb-diagram}

\usepackage{hyperref}

\allowdisplaybreaks

\newcommand{\m}{{-}}
\newcommand{\p}{{+}}

\newcommand{\qmV}{\mathsf{V}}  

\newcommand{\bbeta}{\boldsymbol{\beta}}
\newcommand{\ggamma}{\boldsymbol{\gamma}}

\def\zero{{\mathbf{\bar0}}}
\def\one{{\mathbf{\bar1}}}

\def\aG{\mathfrak{g}}
\def\aN{\mathfrak{n}}
\def\hG{\widehat{\mathfrak{g}}}
\def\WW{\mathscr{W}}
\def\uU{\mathscr{U}}

\newcommand{\A}{\alpha}

\def\betaW{a}
\def\gammaW{a^*}

\def\Vir{\mathop{\mathrm{Vir}}}

\newcommand{\Vac}{\mathop{\mathsf{Vac}}}

\def\N#1{N\!=\!#1}
\def\sS{\boldsymbol{S}}

\def\hD{\widehat{D}(2|1;\alpha)}
\def\D{D(2|1;\alpha)}
\newcommand{\oC}{\mathbb{C}}
\newcommand{\oN}{\mathbb{N}}

\newcommand{\oZ}{\mathbb{Z}}

\def\thalf{\tfrac{1}{2}}
\def\half{\frac{1}{2}}
\def\Dphi{\d\vphi}
\newcommand{\SL}[1]{s\ell(#1)}
\newcommand{\hSL}[1]{\widehat{s\ell}(#1)}
\newcommand{\SSL}[2]{s\ell(#1|#2)}
\newcommand{\hSSL}[2]{\widehat{s\ell}(#1|#2)}

\newtheorem{Thm}{Theorem}[section]
\newtheorem{Lemma}[Thm]{Lemma}

\theoremstyle{definition}

\newtheorem{Rem}[Thm]{Remark}
{\noindent{\nolinebreak\hfill\mbox{\rule{.5em}{.5em}}\,}\par\medskip}

\def\cA{\mathcal{A}} \def\cB{\mathcal{B}} \def\cC{\mathcal{C}}
  
 \def\cH{\mathscr{H}} 
\def\cJ{\mathcal{J}}  
\def\cM{\mathcal{M}} \def\cN{\mathcal{N}} 
 \def\cQ{\mathcal{Q}} 
\def\cT{\mathcal{T}}

\def\bar{\overline}

\def\tensor{\otimes}

\renewcommand{\d}{\partial}

\def\vphi{\varphi}

\def\MPLA{Mod.\ Phys.\ Lett.\ A}

\numberwithin{equation}{section}
\makeatletter 
\@addtoreset{equation}{section}
\def\@secnumfont{\bfseries}
\def\paragraph{\@startsection{paragraph}{4}%
  \z@\z@{-\fontdimen2\font}%
  \normalfont\bfseries}
\def\subparagraph{\@startsection{subparagraph}{5}%
  \z@\z@{-\fontdimen2\font}%
  \normalfont\bfseries}
\makeatother

\begin{document}
\addtolength{\baselineskip}{4pt}
\raggedbottom

\title[Hamiltonian Reduction~$\smash{\widehat{D}(2|1;\alpha)\to}
\widehat{s\ell}(2)\oplus\widehat{s\ell}(2)/\widehat{s\ell}(2)$]{
  \vspace*{-3\baselineskip}\hfill
  {\lowercase{\tt hep-th/0102078}}\\[12pt] The
  $\widehat{s\ell}(2)\oplus\widehat{s\ell}(2)/\widehat{s\ell}(2)$
  Coset Theory as a Hamiltonian Reduction
  of~$\widehat{D}(2|1;\alpha)$}

\author[Feigin]{B.~L.~Feigin} \address{Landau Institute for
  Theoretical Physics, Russian Academy of Sciences}

\author[Semikhatov]{A.~M.~Semikhatov} \address{Tamm Theory
  Division, Lebedev Physics Institute, Russian Academy of Sciences} 

\begin{abstract}
  We show that the coset $\hSL2_{k_1}\oplus\hSL2_{k_2}/\hSL2_{k_1 +
    k_2}$ is a quantum Hamiltonian reduction of the exceptional affine
  Lie superalgebra~$\hD$ and that the corresponding $W$ algebra is the
  commutant of the~$\uU_q\D$ quantum group.
\end{abstract}

\maketitle

\vspace{-2\baselineskip}

\begin{multicols}{2}\addtolength{\parskip}{-1pt}
  \noindent
  {\footnotesize \tableofcontents}
\end{multicols}
\thispagestyle{empty}

\section{Introduction} \label{sec:introduction}
Extensions of bosonic algebras via vertex operators have been seen to
demonstrate a remarkable appearance of affine Lie
\textit{super}algebras: two $\hSL2$ algebras with the levels $k$ and
$k'$ related by $(k+1)(k'+1)=1$ are extended via their spin-$\half$
vertex operators to the affine superalgebra~$\hSSL21$~\cite{[BFST]}.
Moreover, a class of representations of the exceptional affine Lie
superalgebra $\hD$~\cite{[Kac77],[Cornwell]} can also be realized by
extending $\hSL2_k\oplus\hSL2_{k'}$ by \hbox{spin-$\half$} vertex
operators~\cite{[BFST]}.  The emergence of the $\hD$ algebra is rather
intriguing, and this suggests looking for other occurrences of $\hD$
or related algebras hidden behind some known conformal field theory
structures.

\addtolength{\parskip}{2pt}

\bigskip

\noindent
\textbf{1.1. Formulation of the main result.}  We show that $\hD$ is
related to a well-known object, the coset conformal theories
$\hSL2\oplus\hSL2/\hSL2$, via the (quantum) Hamiltonian reduction,
\begin{equation}\label{finding1}
  \text{Quantum Hamiltonian Reduction}(\hD_\varkappa) =
  \frac{\hSL2_{k_1}\oplus\hSL2_{k_2}}{\hSL2_{k_1 + k_2}},
\end{equation}
with the $\alpha$ parameter of the algebra and its level $\varkappa$
expressed as
\begin{equation}\label{the-relations}
  \alpha=-1-\frac{k_1+2}{k_2+2},\qquad
  \varkappa=\frac{-1}{k_1 + k_2 + 4}.
\end{equation}
We also show that
\begin{equation}\label{finding2}
  \smash[t]{\frac{\hSL2_{k_1}\oplus\hSL2_{k_2}}{\hSL2_{k_1 + k_2}}}
  =\WW\!\D,
\end{equation}
where $\WW\!\D$ is the $W$ algebra determined by the root system of
the Lie superalgebra~$\D$; the coset theory on the left-hand side can
be defined as the BRST cohomology of the complex associated with the
$\hSL2_{-4}$ algebra diagonally embedded in
$\hSL2_{k_1}\oplus\hSL2_{k_2}\oplus\hSL2_{k_3}$, where $k_1 + k_2 +
k_3 = -4$.  The central charge is therefore given by
\begin{equation}\label{the-c}
  c = \frac{3k_1}{k_1 + 2} + \frac{3k_2}{k_2 + 2} -
  \frac{3(k_1 + k_2)}{k_1 + k_2 + 2}.
\end{equation}

Two general remarks are in order. First, the algebra $\hD$ admits
\textit{different} Hamiltonian reductions, depending on the chosen
maximal nilpotent subalgebra.  Relations~\eqref{the-relations} apply
to the case where the three simple roots are chosen
fermionic.\footnote{We only consider Hamiltonian reductions that are
  ``\textit{maximal}'' in that all the nilpotent subalgebra currents
  are constrained.  A partial Hamiltonian reduction of $\hD$ leads to
  (nonlinear) $\N4$ superconformal algebras~\cite{[IM],[IMP]}, which
  raises an interesting question regarding a ``secondary'' Hamiltonian
  reduction~\cite{[MR]} from the $\N4$ superconformal algebra to the
  coset.}
Second, the mapping between $(k_1, k_2)$ and $(k,\alpha)$ is not
uniquely fixed because on the one hand, the $\alpha$ parameter is
defined modulo an order-$6$ group of discrete transformations 
and on the other hand, the BRST construction of the coset is invariant
under transpositions of the three levels $k_1$, $k_2$, and $k_3 = -k_1
- k_2 - 4$; this also applies to writing the coset in the GKO
form~\cite{[GKO]}, for example, the left-hand side of~\eqref{finding2}
can be replaced with $\hSL2_{k_2}\oplus\hSL2_{-k_1 - k_2 -
  4}/\hSL2_{-k_1 - 4}$.

Although the notation $\WW\!\D$ for the $W$ algebra explicitly
indicates the root system that determines this algebra, it does not
specify the central charge; we sometimes use the notation
$\WW\!D_{2|1}(k_1,k_2)$ for this $W$ algebra with the central charge
in Eq.~\eqref{the-c}.  Similarly, the level $\varkappa$ in the
notation $\hD_\varkappa$ is conventionally taken to be the level of
one of the $\hSL2$ subalgebras in~$\hD$; a more convenient way to fix
both the $\alpha$ parameter and the level is to specify the levels
$\varkappa_1$, $\varkappa_2$, and $\varkappa_3$ of three $\hSL2$
subalgebras of $\hD$ (these levels are related by
$\frac{1}{\varkappa_1} + \frac{1}{\varkappa_2} + \frac{1}{\varkappa_3}
= 0$); in terms of $k_1$ and $k_2$, we then have
\begin{equation}
  \smash[t]{\varkappa_1 = \frac{1}{k_1+2},\qquad
    \varkappa_2 = \frac{1}{k_2+2},\qquad
    \varkappa_3 = \frac{-1}{k_1+k_2+4}}.
\end{equation}

\medskip

The strategy to arrive at the results in
Eqs.~\eqref{finding1}--\eqref{finding2} involves a combination of
several methods, which are outlined in what follows.

\bigskip

\noindent
\textbf{1.2. Quantum groups and Hamiltonian reduction.} Quantum groups
play an important role in the theory of vertex operator algebras,
similar to the role of a symmetry group.  This has various
manifestations, one of these being the Kazhdan--Lusztig
correspondence~\cite{[KL]} between representation categories of a
vertex operator algebra and of the corresponding quantum group (in
particular cases, equivalence of quasitensor categories has been
proved).  Another important collection of examples is provided by
\textit{free-field realizations}, where the role of the appropriate
quantum group is reminiscent of the Galois group: the quantum group
can be thought of as acting by symmetries on the free-field space such
that the invariants of this action make up a vertex operator algebra.
In practice, the nilpotent (upper-triangular) quantum group generators
are the \textit{screening operators}; in other words, the vertex
operator algebra generators are singled out by the condition that they
commute with the screenings.  This vertex operator algebra is said to
be the \textit{commutant} of the quantum group in the free-field space
(the commutant is always understood as a subalgebra in the algebra of
vacuum descendants of the free-field theory).

Vertex operator algebras defined as the commutant of a chosen set of
screening operators are $W$ algebras.  The general pattern emerging
from a number of known examples is that interesting $W$ algebras
typically arise when the screenings satisfy some special relations.
For generic operators taken as screenings, the corresponding commutant
is trivial; the condition for Virasoro generators to exist in the
commutant restricts the screenings, and the existence of a larger
algebra requires more relations to be satisfied by the screenings,
\hbox{which makes the corresponding quantum group
  ``smaller}.''\footnote{A precise criterion for the occurrence of an
  nontrivial $W$ algebra is not known, however; for example, it is
  \textit{not} true that the corresponding quantum group must
  necessarily be the $q$-deformation of a finite-dimensional
  semisimple Lie algebra.}

A popular class of $W$ algebras are associated with root systems of
semisimple finite-dimensional Lie algebras.  The screening operators
$\sigma_i=\oint e^{\Vec a_i\cdot\Vec\varphi}$ are then constructed
from a set of free fields $\Vec\varphi$ by taking simple root
vectors~$\Vec\alpha_i$ and rescaling them into $\Vec
a_i=\frac{1}{\kappa}\Vec\alpha_i$ (with a parameter related to the
central charge of the $W$~algebra resulting from the reduction).
These $W$ algebras can also be obtained via \textit{Hamiltonian
  reduction}.\footnote{by which we mean the quantum Drinfeld--Sokolov
  reduction---the \textit{maximum} Hamiltonian reduction ``with
  characters,'' i.e., such that the reduction constraints imposed on
  simple root operators are given by $e_i(z) - \chi(e_i(z))=0$ with a
  character $\chi:\widehat\aN\to\oC$.}  Given a system of screening
operators corresponding to the (upper-triangular) Chevalley generators
of a quantum group and then taking the corresponding Lie algebra
(assumed to be finite-dimensional and semisimple) and constructing
its affinisation, one expects to arrive at the algebra whose
Hamiltonian reduction is the commutant of the screenings.  This can be
schematically represented as
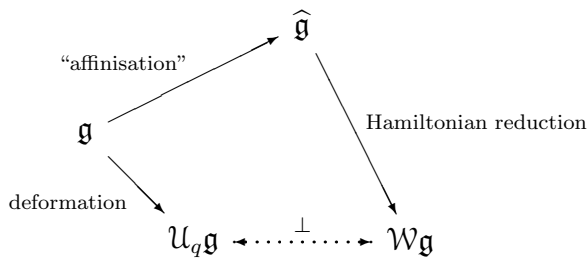
\begin{equation*}
  \begin{diagram}
    \node[3]{\widehat\aG}\arrow{sse,t}{\text{Hamiltonian reduction}}\\
    \node{\aG}\arrow{ene,t,3,}{\text{``affinisation''}}
    \arrow{se,b}{\text{deformation}}\\
    \node[2]{\uU_q\aG}\arrow[2]{e,t,..}{\perp}
    \node[2]{\WW\aG}\arrow[2]{w,b,..}{}
  \end{diagram}
\end{equation*}
where `$\perp$' indicates that each of the two objects is the
commutant of the other in the appropriate free-field space.
Conversely, the Hamiltonian reduction of an affine Lie algebra~$\hG$
(with a semisimple finite-dimensional~$\aG$) is generally expected to
be the commutant of the quantum group $\uU_q\aG$.

Formulating the general scheme outlined in the above diagram as a
theorem would face several subtle points.  The quantum group is only
sensitive to the \textit{exponentials} of the scalar products of the
momenta $\Vec a_i$ used in constructing the screening operators, but
the commutant of the screenings can depend on the actual scalar
products; this raises the problem of finding a ``preferred'' matrix of
scalar products.  For bosonic algebras (i.e., in the cases where $\aG$
is \textit{not} a Lie \textit{super}algebra), the recipe that is known
to work amounts to taking the scalar products of the momenta to be
precisely (up to a common factor) the scalar products of the
corresponding root vectors.  This is not necessarily so for
superalgebras, and the general reformulation of the above scheme is
not known in that case.  In fact, even the definition of the $W$
algebra $\WW\aG$ determined by the root system of $\aG$ requires a
clarification in the case where $\aG$ is a Lie superalgebra.  The
proposal given in what follows applies to the case where the odd roots
are isotropic.

\bigskip

\noindent
\textbf{1.3. Odd roots, fermionic screenings, and $\uU_q\D$.}  For a
Lie superalgebra $\aG$ all of whose odd roots are isotropic, we extend
the notation $\WW\aG$ to denote the algebra that is defined as in the
bosonic case but with the Chevalley operators corresponding to each
odd isotropic root $\alpha_i$ (i.e.,
$\Vec\alpha_i\cdot\Vec\alpha_i=0$) replaced with $\oint e^{\Vec
  a_i\cdot\Vec\varphi}$ where $\Vec a_i$ are determined by $\Vec
a_i\cdot\Vec a_i=1$ and by the condition that the set of all
screenings satisfy the nilpotent subalgebra of~$\uU_q\aG$.

Operators $\oint e^{\Vec a_i\cdot\Vec\varphi}$ with $\Vec a_i\cdot\Vec
a_i=1$ are called \textit{fermionic screenings}; all other screenings
are indiscriminately called \textit{bosonic}.

Examples of systems with fermionic screenings are provided by
two-boson realizations, where two fermionic screenings determine the
nilpotent subalgebra of $\uU_q\SSL21$; its commutant is a nontrivial
$W$ algebra, which also is the Hamiltonian reduction of $\hSSL21$.
(This justifies the notation $\WW\SSL21$ for this $W$ algebra, even
though the condition $\Vec a\cdot\Vec a=1$ is not read off from the
Cartan matrix of~$\SSL21$; this $W$ algebra is also isomorphic to
$\hSL2/u(1)$).  For a system of one fermionic and one bosonic
screening, the resulting $W$ algebra is isomorphic to $\WW\SSL21$
constructed using two fermionic screenings.

In this paper, we consider the case with three fermionic screenings
$(\sigma_1, \sigma_2, \sigma_3)$.  With a single relation imposed on
these operators in the grade $(1,1,1)$ with respect to $(\sigma_1,
\sigma_2, \sigma_3)$, \textit{the three fermionic screenings generate
  the nilpotent subalgebra of the quantum group~$\uU_q\D$}.  This is
the origin of $\D$-related algebras in this paper.

The $\uU_q\D$-relation imposed on the fermionic screenings
translates into a relation on the scalar products of the momenta of
the screenings,
\begin{equation*}
  \Vec{a}_1 \cdot \Vec{a}_2 + \Vec{a}_1 \cdot \Vec{a}_3 + \Vec{a}_2
  \cdot \Vec{a}_3 = n \in\oZ,
\end{equation*}
which involves an arbitrary integer.  We analyze the commutant
of~$(\sigma_1, \sigma_2, \sigma_3)$ for the existence of
higher-dimension operators, in addition to the Virasoro generators.
The first of these operators can occur at dimension~$4$.\pagebreak[3]
We find that for generic values of the other parameters, a primary
\hbox{dimension-4} field exists in the commutant if and only
if~$n=-1$.  We conjecture that for generic values of the other
parameters, a nontrivial $W$ algebra exists in the commutant of
$(\sigma_1, \sigma_2, \sigma_3)$ if and only if~$n=-1$ (the
\textit{conjecture} consists in the ``only if'' part, the converse is
a part of what is proved below).  We define the $W$~algebra $\WW\!\D$
to be the commutant of the three fermionic screenings that represent
the nilpotent subalgebra of~$\uU_q\D$ with~$n=-1$ in the above
equation for the scalar products in the three-boson space.

To mention another example of the occurrence of a fermionic screening
in the three-boson case, we note that the system of one fermionic and
two bosonic screenings, with the bosonic screenings commuting with
each other, gives the nilpotent subalgebra of $\uU_q\D$ corresponding
to a different choice of simple roots.  The commutant is the same $W$
algebra~$\WW\!\D$ as determined by the three fermionic screenings.

The $\WW\!\D$ algebra is similar to other known $W$ algebras in that
it allows \textit{two} systems of screening operators generating
different quantum groups that are Langlands-dual to each other (and
are therefore different in the non-simply-laced case).  For $\WW\!\D$,
these quantum groups are $\uU_q\D$ and
$\uU_{q_1}\SL2\tensor\uU_{q_2}\SL2\tensor\uU_{q_3}\SL2$.  The latter
is represented by three commuting bosonic screenings that play an
important role in establishing the relation of $\WW\!\D$ to the coset
$\hSL2_{k_1}\oplus\hSL2_{k_2}/\hSL2_{k_1 + k_2}$.

\bigskip

\noindent
\textbf{1.4. Vertex-operator extensions of vertex operator algebras.}
To show that $\WW\!\D$ is the coset
$\smash{\hSL2_{k_1}\oplus\hSL2_{k_2}/\hSL2_{k_1 + k_2}}$, we
reconstruct the two $\hSL2_{k_i}$ algebras in the numerator from the
$\smash[t]{\hSL2_{k_1 + k_2}}$ algebra in the denominator and
the~$\WW\!\D$ algebra.  This is done by constructing the
\textit{vertex-operator extension}
\begin{equation}\label{the-extension-0}
  \smash{\WW\!\D\tensor\uU\hSL2_{k_1+k_2}\longrightarrow
  \uU\hSL2_{k_1}\tensor\uU\hSL2_{k_2}}
\end{equation}
involving vertex operators that carry representations of the three
$\uU_q\SL2$ quantum groups.

Vertex-operator extensions highlight the use of quantum groups in
describing \textit{monodromy properties} of vertex operators;
extensions with the help of vertex operators as, e.g.,
in~\eqref{the-extension-0}, require local operators, and these are
usually constructed by taking products of operators carrying dual
quantum group representations and then taking the ``quantum'' trace so
as to obtain operators with trivial monodromies with respect to each
other.

A given set of screening operators can be used to define, in addition
to the commutant $W$ algebra, a natural set of vertex operators of
this $W$ algebra (and their descendants) by selecting all those
free-field operators that generate \textit{finite-dimensional} quantum
group representations under the action of the screenings.  These
vertex operators are therefore labeled by representations of the
quantum group(s) generated by the screenings.
Vectors in the vacuum representation of the $W$ algebra---or
equivalently, the local fields---are then in a $1:1$ correspondence
with quantum-group singlets.

We use vertex-operator extensions to construct the
$\hSL2_{k_1}\oplus\hSL2_{k_2}$ currents\pagebreak[3] and also the
corresponding vertex operators (in all cases, we only use the
$\WW\!\D$ vertex operators that are singlets with respect
to~$\uU_q\D$).  Constructing the $\hSL2_{k_i}$ vertex operators is
based on the ``contraction'' given by the quantum trace in the product
of the two-dimensional $\SL2$ quantum group representations $\oC^2_q$
and $\oC^2_{q'}$ with the ``dual'' quantum group parameters $q=e^{2\pi
  i/(k+2)}$ and $q'=e^{2\pi i/(k'+2)}$.  The idea of this ``duality,''
with $\frac{1}{k+2} + \frac{1}{k'+2}\in\oN$, was discussed
in~\cite{[BFST]} (which was devoted to the case where $\frac{1}{k+2} +
\frac{1}{k'+2}=1$) and is similar to the duality used in
$\mathrm{matter}+\mathrm{gravity}$ theory and also has a counterpart
in solvable lattice models of statistical mechanics, where expressing
the $\mathbf{T}$ operators as the trace of $\mathbf{L}$ operators is
parallel to the above contraction of vertex operators.  In the present
case, we have $\tfrac{1}{k+2} + \tfrac{1}{k'+2} = 0$, which allows us
to take the quantum trace of the product of $\hSL2_{k_1+k_2}$
and~$\WW\!\D$ vertex operators.  This gives the spin-$\half$ vertex
operators for each of the $\hSL2_{k_i}$ algebras (the quantum groups
that are not involved in the contraction become quantum group
symmetries of the resulting vertex operators); the spin-$\half$
operators generate the entire algebra of vertex operators.  The
$\hSL2_{k_i}$ currents can be reconstructed via a similar
vertex-operator extension involving a contraction in the product of
three-dimensional quantum group representations (such that the result
is a singlet with respect to all quantum groups).

\bigskip

\noindent
\textbf{1.5. Hamiltonian reduction of $\hD$.} We perform the
Hamiltonian reduction of $\hD$ with the help of a ``BRST'' operator
$\cQ$ implementing the constraints imposed on the nilpotent subalgebra
currents.  The reduction is such that the cohomology of $\cQ$
certainly contains a Heisenberg algebra $\cH_0$, which is a
``trivial'' piece guaranteed by the nature of the reduction, while the
nontrivial problem is to find a $W$ algebra commuting with~$\cH_0$.
The Virasoro generators of this $W$~algebra can be constructed
explicitly, and it is then verified that their central charge is the
one in Eq.~\eqref{the-c}.  To show that the cohomology contains the
entire $\WW\!\D$ algebra, we introduce a filtration on the BRST
complex such that the BRST operator splits as $\cQ=\cQ^{(0)} +
\cQ^{(1)} + \cQ^{(2)}$ with $\cQ^{(i)}$ decreasing the filtration
index by~$i$.  The cohomology of $\cQ^{(0)}$ is given by (apart from
the ``trivial'' Heisenberg algebra) a Heisenberg algebra represented
by three scalar fields.  On this algebra, the action of $\cQ^{(1)}$
amounts to the action of three fermionic screenings that satisfy the
nilpotent subalgebra of $\uU_q\D$ (and the next differential acts
trivially).  This allows us to show that the Hamiltonian reduction of
$\hD$ is the $W$ algebra~$\WW\!\D$.

\bigskip

In Sec.~\ref{sec:varphi123}, we start with three fermionic screenings
in a three-boson realization and impose one relation in the grade
$(1,1,1)$ with respect to these operators.  This makes three generic
fermionic screenings into those representing the nilpotent subalgebra
of~$\uU_q\D$.  In Sec.~\ref{sec:commutant}, we explicitly find the
lowest-dimension $\WW\!\D$ operators.  In Sec.~\ref{sec:R-scr}, we
construct the screenings representing the nilpotent subalgebra of the
Langlands-dual quantum group (the sum of three $\uU_q\SL2$).  In
Sec.~\ref{sec:constr-coset-deform}, we arrive at the same system of
three $\uU_q\SL2$ screenings and the $W$ algebra in their commutant by
``deforming'' the $\hSL2$ WZW theory.  This allows us to find two
remarkable vertex operators for $\WW\!\D$ that are analogues of the
$\Phi_{21}$ operators for the Virasoro algebra and also a triplet
operator.  These $\WW\!\D$ vertex operators are used in
Sec.~\ref{sec:reconstruct} to reconstruct the $\hSL2_{k_i}$ vertex
operators and currents and thus to show that $\WW\!\D$ is indeed the
algebra of the $\hSL2_{k_1}\oplus\hSL2_{k_2}/\hSL2_{k_1 + k_2}$ coset.
In Sec.~\ref{sec:qgs-and-vertices}, we review several relevant points
pertaining to vertex-operator extensions, in
Sec.~\ref{sec:coset-recall} recall defining properties of the coset,
and use these in Sec.~\ref{sec:reconstruct-currents} to reconstruct
the $\hSL2_{k_1}\oplus\hSL2_{k_2}$ currents (and in
Sec.~\ref{subsec:reconstruct}, the corresponding vertex operators) via
a vertex-operator extension.  The Hamiltonian reduction of $\hD$
leading to this $W$ algebra is worked out in Sec.~\ref{sec:Ham-red}.
A number of related issues and further directions are discussed in
Sec.~\ref{sec:discussion}.  Appnedix~\eqref{app:formula} contains an
explicit formula for the dimension-$4$ operator in the commutant of
the screenings.  In Appendix~\ref{app:D-all}, we give our conventions
on $\D$ and the $\hD$ commutation relations.

\section{Fermionic screenings for $\WW\!\D$} \label{sec:varphi123} In
this section, we show that the nilpotent subalgebra of $\uU_q\D$ can
be generated by three fermionic screenings in the three-boson
realization such that the commutant of these screenings is a
nontrivial $W$ algebra.

We recall (see~\cite{[Kac77],[Cornwell]}) that $\D$ is the
superalgebra with the bosonic
part~$\D_{\zero}=\SL2\oplus\SL2\oplus\SL2$ and the action
of~$\D_{\zero}$ on $\D_{\one}$ given by the product of the
two-dimensional representations.  Commutation relations of~$\D$ are
written in Appendix~\ref{app:D-all}, but in this section, we only use
the fact that $\D$ admits a simple root system where all the three
roots are fermionic; the Chevalley generators $\psi_i$, $i=1,2,3$,
then satisfy $[\psi_1, \psi_1]=0$, $[\psi_2, \psi_2]=0$, $[\psi_3,
\psi_3]=0$ (where $[~,~]$ denotes the \textit{super}commutator), and
\begin{equation}\label{relation-L}
  [\psi_2, [\psi_1, \psi_3]] +
  (\alpha+1)[\psi_3, [\psi_1, \psi_2]] = 0
\end{equation}
(a non-Serre-type relation in the approach of~\cite{[Leites]}).  Thus,
the nilpotent subalgebra contains three even elements $[\psi_1,
\psi_2]$, $[\psi_2, \psi_3]$, and $[\psi_3, \psi_1]$ and one more odd
element, which up to proportionality is given by any of the triple
commutators.  The algebra admits an invariant form and can therefore
be quantized.  Remarkably, the crucial relation in $\uU_q\D$---the
``quantum'' analogue of~\eqref{relation-L}---follows from a simple
algebraic problem involving three screening operators.

\subsection{A cubic relation on three fermionic screenings}
\label{subsec:cubic} We consider a formal algebraic problem capturing
the commutation properties of screening operators for the $\D$ root
system.  We construct three screenings as
\begin{equation}\label{sigma-i}
  \sigma_i = \int S_i(z),\qquad i = 1, 2, 3,
\end{equation}
where $S_i$ satisfy formal relations
\begin{equation}\label{SSq}
  S_i(z) S_j(w) = q_{ij}\,S_j(w) S_i(z),\qquad z > w.
\end{equation}
In the case where $q_{ij}=q^{a_{ij}}$, with $a_{ij}$ being the matrix
of scalar products of a simple root system (for the respective range
of $i,j$),\pagebreak[3] the corresponding screenings satisfy the
relations of the nilpotent subalgebra of the quantum group determined
by the chosen root system.

For $\sigma_i$ to generate the nilpotent subalgebra of $\uU_q\D$
corresponding to three fermionic roots, we take each $S_i(z)$ to be a
fermionic operator and consider the simplest nontrivial relation that
can be imposed on all the three operators~$\sigma_i$.  The first
nontrivial relation is in the grade $(1,1,1)$ with respect to
$(\sigma_1,\sigma_2,\sigma_3)$.
The only invariant relation in this grade states that the six
operators given by the triple integrals of
$S_{p(1)}(z)S_{p(2)}(w)S_{p(3)}(u)$ with $z<w<u$, where $p$ runs over
permutations on three elements, are linearly dependent (but no
stronger conditions hold).  Using~\eqref{SSq}, each of the six
operators can be expressed through six ``basis'' operators
$S_1(z)S_2(w)S_3(u)$ in the respective domains
\begin{alignat*}{3}
  &z<w<u,& \quad &w<u<z,& \quad &u<z<w,\\*
  &z<u<w,& &w<z<u, & &u<w<z.
\end{alignat*}
The trilinear relation amounts to the condition that the $6\times6$
matrix relating the two sets of operators must have rank~$5$.  This
matrix
\begin{equation}\label{the-matrix}
  \begin{pmatrix}
    1&q_{23}&q_{12}& q_{12}q_{13}& q_{13}q_{23}&
    q_{12}q_{13}q_{23}\\
    q_{23}&1&q_{12}q_{23}& q_{12}q_{13}q_{23}&
    q_{13}&q_{12}q_{13}\\
    q_{12}&q_{12}q_{23}&1& q_{13}&q_{13}q_{12}
    q_{23}&q_{13}q_{23}\\
    q_{12}q_{13}& q_{12}q_{23}q_{13}& q_{13}&1&q_{12}q_{23}&
    q_{23}\\
    q_{13}q_{23}&q_{13}& q_{12}q_{13}q_{23}&
    q_{12}q_{23}&1&q_{12}\\
    q_{12}q_{13}q_{23}& q_{12}q_{13}& q_{13}q_{23}&q_{23}& q_{12}&1
  \end{pmatrix}
\end{equation}
has the determinant $(1-q_{12}^2)^2 (1-q_{13}^2)^2 (1-q_{23}^2)^2 (1 -
q_{12}^2 q_{13}^2 q_{23}^2)$, and we arrive at the
condition\footnote{The other vanishings are with the multiplicity~$2$.
  We do not consider these ``strongly degenerate'' cases here, because
  we are interested in $\D$-related structures, but they can also be
  interesting.  For example, $q_{ij}=-1$ means that the two fermionic
  screenings point in the opposite directions; the $\hSL2$ algebra
  determined by such a pair of fermionic screenings then has the
  level~$-2$, which makes a story of its own.}
\begin{equation}\label{q-cond}
  \smash{q_{12}^2\,q_{13}^2\,q_{23}^2} = 1.
\end{equation}
The relation on $\sigma_i$ is then given by
\begin{multline}\label{relation-q}
  (q_{13}^2 - 1) \sigma_1\sigma_2\sigma_3 + \tfrac{1 - q_{13}^2
    q_{23}^2}{q_{23}} \sigma_1\sigma_3\sigma_2 - q_{12} q_{13}^2 (1 -
  q_{23}^2) \sigma_2\sigma_1\sigma_3 +{}\\
  {} + q_{12}q_{13} (1 - q_{13}^2 q_{23}^2) \sigma_2\sigma_3\sigma_1 -
  \tfrac{q_{13} (1 - q_{23}^2)}{q_{23}} \sigma_3\sigma_1\sigma_2 +
  q_{12}q_{13} (q_{13}^2 - 1) q_{23} \sigma_3\sigma_2\sigma_1 =0.
\end{multline}

We now realize the $S_i$ operators in terms of three free fields
$\Vec\varphi=\{\varphi_1,\varphi_2,\varphi_3\}$ with the operator
products
\begin{equation}
  \d\vphi_\alpha(z)\,\d\vphi_\beta(w) =
  \smash[t]{\frac{\delta_{\alpha\beta}}{(z-w)^2}}.
\end{equation}
Introducing three 3-dimensional vectors $\Vec{a}_i$, we set (with the
Cartesian scalar product)
\begin{equation}\label{S-i}
  S_i = \smash{e^{\Vec{a}_i\cdot\Vec{\vphi}}}.
\end{equation}
The monodromies in~\eqref{SSq} are then given by
\begin{equation}\label{express-q}
  q_{ij}= \smash{e^{\pi i \Vec a_i\cdot\Vec a_j}}.
\end{equation}
The trilinear relation ensured by~\eqref{q-cond} now reformulates as
the condition that the sum of the scalar products be an integer,
\begin{equation}\label{conditions-a}
  \Vec{a}_1 \cdot \Vec{a}_2 + \Vec{a}_1 \cdot \Vec{a}_3 + \Vec{a}_2
  \cdot \Vec{a}_3 = n \in\oZ.
\end{equation}
In accordance with the recipe to construct fermionic screenings, we
also have the conditions
\begin{equation}\label{Ai-Aj}
  \Vec{a}_i\cdot\Vec{a}_i=1,\qquad i=1,2,3.
\end{equation}
To solve the four equations in~\eqref{conditions-a} and~\eqref{Ai-Aj}
for $\Vec{a}_i$, we parametrize the scalar products as
\begin{align}
  \Vec{a}_1\cdot\Vec{a}_2 ={}& k_2+1,\\
  \Vec{a}_1\cdot\Vec{a}_3 ={}& k_1+1,\\
  \Vec{a}_2\cdot\Vec{a}_3 ={}& n - 2 - k_1 - k_2
\end{align}
in terms of arbitrary parameters $k_1$ and $k_2$.  The general
solution of~Eqs.~\eqref{conditions-a}--\eqref{Ai-Aj} modulo
reflections and three-dimensional rotations is then given
by\footnote{The imaginary unit means that the signature of the
  three-dimensional space of the scalars is $(+, -, +)$, i.e.,
  $\varphi_2$ could be redefined such that
  $\d\varphi_2(z)\d\varphi_2(w)=-1/(z-w)^2$ and $i$ consistently
  removed.}
\begin{align}\label{a-found1}
  \Vec{a}_1 ={}& \{1, 0, 0\},\\
  \Vec{a}_2 ={}& \{k_2 + 1, -i \sqrt{k_2 (k_2+2)}, 0\},\\
  \Vec{a}_3 ={}& \{k_1 + 1, -i \tfrac{2 k_2+k_1 (k_2+2)-n+3}{\sqrt{k_2
      (k_2+2)}}, \sqrt{\tfrac{2(k_2 + 2) k_1^2 + 2 (k_2 + 2) (k_2 - n
      + 3) k_1 + (2 k_2 - n + 3)^2}{k_2 (k_2+2)}}\,\}.\label{a-found3}
\end{align}
Different values of $n$ correspond to different theories in the
commutant of $(\sigma_1,\sigma_2,\sigma_3)$.  We choose the value
corresponding to~$\WW\!\D$ by studying the commutant.

\subsection{The commutant of the screenings} \label{sec:commutant}
With $S_i=e^{\Vec a_i\cdot\Vec\varphi}$ constructed in
Sec.~\ref{subsec:cubic}, we next look for the commutant of the
$\sigma_i$ operators in the three-boson space.  To begin with
dimension two, there is one such operator, and it is the
energy-momentum tensor
\begin{multline}\label{eq:T}
  T = \thalf \Dphi_1 \Dphi_1 + \thalf \Dphi_2\Dphi_2 + \thalf
  \Dphi_3\Dphi_3 - \thalf\d^2\varphi_1 +
  \frac{i}{2}\sqrt{\frac{k_2}{k_2 + 2}}\d^2\varphi_2\\
  {} +
  \half\sqrt{\frac{k_2}{k_2+2}}\,\frac{(2 k_2-n+3)}{
    \sqrt{2 (k_2+2) k_1^2+2 (k_2+2) (k_2-n+3) k_1+ (2
      k_2-n+3)^2}}\d^2\varphi_3. 
\end{multline}
Its central charge is given by
\begin{equation}
  c = \frac{3 k_1 k_2 (k_1+k_2-n+3)}{
    2 (k_2+2) k_1^2 +2 (k_2+2) (k_2-n+3) k_1+(2 k_2-n+3)^2}.
\end{equation}

Searching for higher-dimension primary operators in the commutant, we
find that a symmetry enhancement occurs for $n=-1$, when the commutant
involves a primary operator with the minimum possible dimension~$>2$,
i.e., dimension~$4$.
\begin{Lemma}
  For generic $k_1$ and $k_2$, an operator of conformal dimension~$4$
  that commutes with the screenings $\sigma_i$ determined by
  Eqs.~\eqref{sigma-i}, \eqref{S-i},
  and~\eqref{conditions-a}--\eqref{Ai-Aj}\pagebreak[3] and is primary
  with respect to the energy-momentum tensor~\eqref{eq:T} exists if
  and only if~$n=-1$.
\end{Lemma}
This is shown by solving the system of equations ensuring that a
general dimension-4 operator commutes with the three fermionic
screenings (there are 51 potentially possible operator terms
constructed out of the free fields, with two terms corresponding to
descendants of the energy-momentum tensor). As a function of~$n$, the
determinant of the system vanishes if and only if $n=-1$ (for generic
$k_1$ and $k_2$), and the corresponding solution gives the operator
written in Appendix~\ref{app:formula}.

We thus consider the $n=-1$ case in the above theory.  The
energy-momentum tensor $T|_{n=-1}$ (which we denote simply by $T$ in
what follows),
\begin{multline}\label{T-become}
  T=\thalf \d\varphi_1 \d\varphi_1 + \thalf \d\varphi_2 \d\varphi_2 +
  \thalf \d\varphi_3 \d\varphi_3 -
  \thalf \d^2\varphi_1 \\*
  {}+ \frac{i\sqrt{k_2}}{2\sqrt{k_2 + 2}} \d^2\varphi_2 -
  \frac{\sqrt{k_2}}{\sqrt{2(k_1 + 2)(k_1 + k_2 + 2)}} \d^2\varphi_3,
\end{multline}
has the central charge given by Eq.~\eqref{the-c}.

\textit{We define the $W$ algebra $\WW\!\D$ to be the commutant of the
  operators $\sigma_i=\oint e^{\Vec a_i\cdot\Vec\varphi}$ with $\Vec
  a_1\cdot\Vec a_2 + \Vec a_1\cdot\Vec a_3 + \Vec a_2\cdot\Vec a_3
  =-1$ and $\Vec a_i\cdot\Vec a_i=1$ that generate the nilpotent
  subalgebra of~$\uU_q\D$}.

For $n=-1$, we now summarize the screenings that we use in what
follows:
\begin{align}\label{sigma-1-final}
  \sigma_1=\oint e^{\Vec a_1\cdot\Vec\varphi},\qquad
  \Vec{a}_1 ={}& \{1, 0, 0\},\\
  \sigma_2=\oint e^{\Vec a_2\cdot\Vec\varphi},\qquad
  \Vec{a}_2 ={}& \{k_2 + 1, -i\sqrt{k_2(k_2 + 2)}, 0\},
  \label{sigma-2-final}\\
  \sigma_3=\oint e^{\Vec a_3\cdot\Vec\varphi},\qquad \Vec{a}_3 ={}&
  \{k_1 + 1, -i\tfrac{(k_1 + 2)\sqrt{k_2 + 2}}{ \sqrt{k_2}},
  \sqrt{\tfrac{2(k_1 + 2)(k_1 + k_2 +
      2)}{k_2}}\}.\label{sigma-3-final}
\end{align}

\begin{Rem} The semiclassical limit of the trilinear relation on 
  $\sigma_i$ is the $\D$ relation~\eqref{relation-L}.  Indeed,
  inserting~\eqref{express-q} with the above $\Vec{a}_i$ in
  Eq.~\eqref{relation-q}, we obtain
  \begin{multline}\label{relation-k}
    (e^{2\pi ik_1} - 1) \sigma_1\sigma_2\sigma_3 - (e^{\pi i
      (k_1 + k_2)} - e^{\pi i( k_1 - k_2 )}) \sigma_1\sigma_3\sigma_2
    + (e^{\pi i (2 k_ 1 + k_2)} - e^{-\pi ik_2})
    \sigma_2\sigma_1\sigma_3 +{}\\{} + (e^{\pi i(k_1 + k_2)} -
    e^{i\pi(k_1 - k_2 )}) \sigma_2\sigma_3\sigma_1 - (e^{\pi
      i(2 k_1 + k_2)} - e^{-\pi ik_2} ) \sigma_3\sigma_1\sigma_2 -
    (e^{2\pi ik_1} - 1)\sigma_3\sigma_2\sigma_1 = 0.
  \end{multline}
   Semiclassically, in the first nontrivial order, this becomes
  \begin{equation} \label{class-relation}
    k_1[[\sigma_1,\sigma_2],\sigma_3] + k_2[\sigma_2,[\sigma_1,
    \sigma_3]] = 0.
  \end{equation}  
  Comparing~\eqref{class-relation} and~\eqref{relation-L}, we conclude
  that in the classical limit, the algebra generated by $\sigma_1$,
  $\sigma_2$, and $\sigma_3$ is the nilpotent subalgebra of $\D$ with
  (using the identifications $\sigma_1 = \psi_1$, $\sigma_2 = \psi_3$,
  and $\sigma_3 = \psi_2$)
  \begin{equation}\label{alpha-semi-classical}
    \alpha = -1 - \frac{k_2}{k_1}.
  \end{equation}
  The ``quantum''
  relation~\eqref{relation-k} is the corresponding quantum-group
  deformation of~\eqref{relation-L}.
\end{Rem}

\subsection{Bosonic screenings} \label{sec:R-scr} Any chosen pair of
the fermionic screenings $\{\sigma_i,\sigma_j\}$, $i\neq j$,
determines (the nilpotent subalgebra of) the quantum group
$\uU_q\SSL21$ (with the simple roots of $\SSL21$ chosen fermionic).
Accordingly, the commutant of these two screenings in the appropriate
two-boson subspace is the $W$ algebra~$\WW\SSL21$.
This algebra commutes with a third, bosonic, screening operator of a
\textit{non-vertex-operator} form (the product of a current and an
exponential). The occurrence of this screening can be seen by invoking
the $\hSL2$ argument: the commutant of any \textit{pair} of fermionic
screenings in the three-boson realization is (the symmetric
realization of) the $\hSL2$ algebra, where the occurrence of the third
screening, of a non-vertex-operator form, is well known.  The
$\WW\!\D$ algebra is the intersection of the three algebras of the
form $(\WW\SSL21 \tensor (\text{Heisenberg}))$.  Therefore, the
commutant of $\WW\!\D$ contains the bosonic screenings
\begin{gather}\label{Rij}
  \rho_{ij}=\oint R_{ij},\qquad
  R_{ij}=J_{ij}\,e^{\Vec r_{ij}\cdot\Vec{\varphi}},
  \quad i\neq j,\\[-6pt]
  \intertext{where}
  \label{alpha-ij}
  \Vec r_{ij} =
  -\smash[t]{\frac{1}{(\Vec{a}_i\cdot\Vec{a}_j)+1}(\Vec{a}_i +
    \Vec{a}_j)}
\end{gather}
and $J_{ij}$ is a linear combination of $\Vec{a}_i\cdot\d\Vec\varphi$
and $\Vec{a}_j\cdot\d\Vec\varphi$.  By adding a total derivative to
$R_{ij}$ and by overall normalizations, the coefficients in this
linear combination can be fixed arbitrarily.

The vectors $\Vec r_{ij}$ satisfy 
\begin{equation}
  (\Vec r_{12})^2=\frac{2}{k_2+2},\quad
  (\Vec r_{13})^2=\frac{2}{k_1+2},\quad
  (\Vec r_{23})^2=\frac{-2}{k_1 + k_2+2}.
\end{equation}
As another consequence of setting $n=-1$ in~\eqref{a-found3}, we have
\begin{equation}\label{eq:2}
  \Vec r_{12}\cdot\Vec r_{13}=0,\qquad
  \Vec r_{12}\cdot\Vec r_{23}=0,\qquad
  \Vec r_{23}\cdot\Vec r_{13}=0
\end{equation}
and (with $\Vec r_{ij}=\Vec r_{ji}$ for $i\neq j$)
\begin{equation}\label{useful}
  \begin{split}
    \Vec a_i\cdot\Vec r_{jk} &= 1, \quad i,j,k~\text{all distinct},\\
    \Vec a_i\cdot\Vec r_{ij} &= -1.
  \end{split}
\end{equation}
The following statement is now readily verified.
\begin{Lemma}
  The operators $\rho_{12}$, $\rho_{13}$, and $\rho_{23}$ pairwise
  commute.  Each $\rho_{ij}$ commutes with $\sigma_1$, $\sigma_2$, and
  $\sigma_3$.
\end{Lemma}

Explicitly, the integrands of the bosonic screenings are given by
\begin{align}
  R_{12}={}&(\alpha_{12} \d\varphi_1+i \beta_{12} \d\varphi_2)
  e^{-\varphi_1 + i\sqrt{\frac{k_2}{k_2+2}}\varphi_2},\\
  R_{13}={}& \Bigl((\alpha_{13} + \beta_{13}(k_1 + 1))\d\varphi_1 - i
  \beta_{13} 
  (k_1 + 2)\sqrt{\tfrac{k_2 + 2}{k_2}}\d\varphi_2\\*[-2pt]
  &{}+ \sqrt{2}
  \beta_{13} \sqrt{\tfrac{(k_1 + 2)(k_1 + k_2 + 2)}{k_2}}
  \d\varphi_3\Bigr) e^{-\varphi_1 + i\sqrt{\frac{k_2+2}{k_2}}\varphi_2
    - \sqrt{\frac{2(k_1 + k_2 + 2)}{k_2(k_1+2)}}\varphi_3},\notag\\
  R_{23}={}& \Bigl((\alpha_{23}(k_2 + 1) + \beta_{23}(k_1 +
  1))\d\varphi_1 - i 
  (\beta_{23}(k_1 +2) + \alpha_{23}k_2) \sqrt{\tfrac{k_2 +
      2}{k_2}}\d\varphi_2 
  \\*[-2pt]
  &{} + \sqrt{2}\beta_{23}\sqrt{\tfrac{(k_1 + 2)(k_1 + k_2 + 2)}{k_2}}
  \d\varphi_3\Bigr) e^{\varphi_1 - i\sqrt{\frac{k_2+2}{k_2}}\varphi_2
    + \sqrt{\frac{2(k_1+2)}{k_2(k_1+k_2+2)}} \varphi_3}.\notag
\end{align}
In the next section, these operators are reproduced from a different
argument with fixed values of the $\alpha_{ij}$ and $\beta_{ij}$
coefficients.

Thus, we have constructed screening operators that generate the
nilpotent subalgebras of the quantum groups $\uU_q\D$ and
$\uU_{q_1}\SL2\tensor\uU_{q_2}\SL2\tensor\uU_{q_3}\SL2$ (with the
quantum-group parameters that are explicitly written in what follows).
In the next section, we construct $\WW\!\D$ vertex operators carrying
representations of the $\uU_{q_i}\SL2$ quantum groups.

\section{$\WW\!\D$ by deformation and the $\WW\!\D$ vertex operators}
\label{sec:constr-coset-deform} In this section, we show that the
$\WW\!\D$ algebra can be constructed by deforming (a subalgebra of)
the $\uU\hSL2$ algebra.\footnote{By~$\uU$, we here mean the vertex
  operator algebra and sometimes, its vacuum representation (this must
  not lead to a confusion).}  This allows us to construct $\WW\!\D$
vertex operators, in particular those that are ${212}$, ${122}$,
${221}$, and ${113}$ representations of the three $\uU_q\SL2$ quantum
groups corresponding to the $\rho_{ij}$ screenings.
The operators that we need in the next sections are summarized
in~\eqref{the-three}.

The outline of this section is as follows.  In
Sec.~\ref{subsec:standard-realisation}, we deform $\uU\hSL2$ into a
$W$ algebra by taking the commutant of two bosonic screenings.
Together with the standard Wakimoto screening, the bosonic screenings
generate three commuting $\uU_q\SL2$ quantum groups.  In
Sec.~\ref{sec:qg-doublets}, we construct vertex operators that are
singlets with respect to one of these quantum groups and are doublets
with respect to the other two.  In Sec.~\ref{sec:sym-bos}, we use the
``\textit{symmetric}'' bosonization, which allows us to additionally
construct the vertex operator that is a triplet with respect to one of
the three $\SL2$ quantum groups.  We then show that the $W$ algebra
obtained by deforming $\uU\hSL2$ is in fact $\WW\!\D$ and explicitly
map the ``deformed'' picture onto that in Sec.~\ref{sec:varphi123}.

\subsection{A ``$\beta$-deformation'' of $\smash{\protect\hSL2}$}
\label{subsec:standard-realisation}  We start with the level-$k$
$\hSL2$ algebra, which in terms of operator products is given by
\begin{equation}
  \begin{split}
    J^+(z)\,J^-(w) &= \frac{k}{(z-w)^2} + \frac{2J^0}{z-w},\\
    J^0(z)\,J^\pm(w) &= \frac{\pm J^\pm}{z-w},\qquad J^0(z)\,J^0(w) =
    \frac{k/2}{(z-w)^2},
  \end{split}
\end{equation}
and consider the subalgebra in $\uU\hSL2$ that commutes with the
\textit{Lie} algebra $\SL2$.  We next deform this vertex operator
algebra using the operators
\begin{equation}\label{sigmapm}
  \sigma^\pm = \oint S^\pm,\qquad
  S^\pm(z) =  J^\pm(z)\,e^{\beta^{\pm}\!\int\limits^{\;z}\! J^0}
\end{equation}
with the two parameters $\beta^+$ and $\beta^-$ related by
\begin{equation}
  \beta^{\mp} = \frac{\beta^{\pm}}{1 \pm \frac{k}{2}\beta^{\pm}},
\end{equation}
which implies that
\begin{equation}
  [\sigma^+,\sigma^-]=0.
\end{equation}
We let $W_*$ denote the commutant of $\sigma^+$ and $\sigma^-$ in
$\uU\hSL2$.

We write $\beta\equiv\beta^+$ and $\beta'\equiv\beta^-$ in what
follows; thus,
\begin{equation}
  \beta' = \frac{\beta}{1 + \frac{k}{2}\beta}.
\end{equation}

We next introduce the Wakimoto bosonization
\begin{align}
  J^+ &= -\betaW,\\
  J^0 &= \betaW\gammaW + \sqrt{\tfrac{k+2}{2}}\,\d\varphi,\\
  J^- &= (\gammaW)^2\betaW + \sqrt{2(k+2)}\,\gammaW \d\varphi + k
  \d\gammaW
\end{align}
and additionally represent the bosonic ghosts $\betaW$ and $\gammaW$
(with the operator product $\betaW(z)\gammaW(w)=\frac{-1}{z-w}$) in
terms of free scalars $f$ and $\phi$ with $\d f(z)\d
f(w)=\frac{1}{(z-w)^2}$ and $\d\phi(z)\d\phi(w)=\frac{-1}{(z-w)^2}$
via
\begin{equation}\label{beta-gamma-bos}
  \betaW = -e^{-f-\phi},\qquad \gammaW=\d f\,e^{f+\phi}.
\end{equation}
This gives a three-boson realization of $\hSL2$ (known as the standard
or the asymmetric realization).  Normal ordering of all composite
operators is understood.  We also recall the Wakimoto bosonization
screening
\begin{equation}\label{Wak-scr}
  \sigma_{\mathrm{W}} = \oint S_{\mathrm{W}},\qquad
  S_{\mathrm{W}}=-\betaW\,e^{-\sqrt{\frac{2}{k+2}}\varphi}.
\end{equation}

The $W_*$ algebra is now selected from the free-field space as the
commutant of $\sigma_{\mathrm{W}}$, $\sigma^+$, and $\sigma^-$ (all of
which are called screenings hereafter).  By a direct calculation, we
obtain
\begin{Lemma}
  The algebra $W_*$ contains a Virasoro algebra
  with the central charge
  \begin{equation}\label{c-perp}
    c_\perp = \frac{3k}{k+2} - \frac{3k\beta^2 (k\beta + 4)^2}{
      2(k\beta + 2)^2}.
  \end{equation}
\end{Lemma}
For the ``undeformed'' value $\beta=0$, this is indeed the $\hSL2$ WZW
model central charge.  In the calculation, we used that the operators
in the integrands of $\sigma^\pm$ rewrite as
\begin{equation}\label{plus-integrand}
  J^+(z)\,e^{\beta\!\int\limits^{\;z}\! J^0}=
  e^{(\beta-1)\phi}\,e^{\beta\sqrt{\frac{k+2}{2}}\varphi}\,e^{-f}(z)
\end{equation}
and
\begin{multline}\label{minus-integrand}
  J^-(z)\,e^{\beta'\!\int\limits^{\;z}\! J^0}= \Bigl((k + 1)\d^2 f +
  (k + 1)\d f \d f +
  (k + 2)\d f \d\phi\\
  {}+ \sqrt{2(k + 2)} \d \varphi \d f\Bigr) e^{(\beta' + 1)\phi}
  e^{\beta'\sqrt{\frac{k + 2}{2}}\varphi} e^{f}(z).
\end{multline}

\subsection{$122$-type quantum group representations}
\label{sec:qg-doublets} We next recall the spin-$\half$ vertex 
operator of $\hSL2_k$ and deform it into a certain vertex operator
for~$W_*$.  The spin-$\half$ vertex is a 2-dimensional quantum group
representation with the highest-weight vector given by
\begin{equation}
  \Psi \equiv \Phi_{\half} =
  \smash{e^{\frac{1}{\sqrt{2(k+2)}}\varphi}}.
\end{equation}
The action of the horizontal $\SL2$ subalgebra on $\Psi$ generates a
two-dimensional representation,
\begin{equation}
  J^-_0 \Psi = \gammaW e^{\frac{1}{\sqrt{2(k+2)}}\varphi},\qquad
  (J^-_0)^2 \Psi = 0,\qquad J^+_0 \Psi = 0.
\end{equation}
These properties persist under the ``deformation'' of $J^\pm$ by
$e^{\beta^{\pm}\!\int\limits^{\;z}\! J^0}$ if the vertex is deformed
into
\begin{equation}
  \Psi_\beta(w) = \Psi(w)\,e^{-\half\beta'\!\int\limits^{\;\;w}\!J^0}
  =e^{-\half\beta'\phi}\,
  e^{\bigl(\frac{1}{\sqrt{2(k+2)}} -
    \frac{\beta'}{2}\sqrt{\frac{k+2}{2}}\bigr)\varphi}(w).
\end{equation}
(we recall that $\beta'$ is expressed through $\beta$ and the
level~$k$).  Namely, we have
\begin{equation}\label{ope-1}
  J^-(z)\,e^{\beta'\!\int\limits^{\;z}\! J^0}\cdot
  \Psi_\beta(w)
  =(z-w)^{(\beta' + 1)\frac{\beta'}{2} +
    \beta'(-\frac{k+2}{4}\beta' 
    + \half) - 1}\,
  \d f\,e^{(\half\beta' + 1)\phi}\,
  e^{\frac{\half\beta'(k+2)+1}{\sqrt{2(k+2)}}\varphi} e^{f}(w).
\end{equation}
Remarkable cancellations of poles occur in acting
with~\eqref{minus-integrand} on the right-hand side of the last
formula, resulting in
\begin{equation}\label{ope-2}
  J^-(z)\,e^{\beta'\!\int\limits^{\;z}\! J^0}\cdot
  \d f\, e^{(\half\beta' + 1)\phi}\,
  e^{\frac{\half\beta'(k+2)+1}{\sqrt{2(k+2)}}\varphi} e^{f}(w)
  \sim
  (z-w)^{-(\beta'+1)(\half\beta'+1) +
    \beta'(\frac{k+2}{4}\beta' + \half)+1}.
\end{equation}

Two effects now occur.  First, $(\sigma^-)^2 \Psi_\beta(w)$ is a local
operator, because the sum of the exponents in~\eqref{ope-1}
and~\eqref{ope-2} is an integer,
\begin{equation}
  [(\beta' + 1)\tfrac{\beta'}{2} +
  \beta'(-\tfrac{k+2}{4}\beta' 
  + \thalf) - 1] +
  [-(\beta' + 1)(\thalf\beta' + 1) +
  \beta'(\tfrac{k+2}{4}\beta' + \thalf)+1]=-1.
\end{equation}
Second, moreover, the relevant first-order pole vanishes (one extra
positive power comes from the integration), and therefore, applying
the standard argument, we conclude that
\begin{equation}
  (\sigma^-)^2 \Psi_\beta(w)=0.
\end{equation}

Thus, $\Psi_\beta$ is a part of a doublet under the action
of~$\sigma^-$.  At the same time, it is a singlet with respect to
$\sigma^+$.  Further, the spin-$\half$ vertex operator is a doublet
under the action of the Wakimoto screening, and because the
deformation involves only the $\hSL2$ current $J^0$, this property
persists for~$\Psi_\beta$.  We thus conclude that $\Psi_\beta$ is the
$(1,2,2)$ representation of (the quantum $\SL2$ groups with the
respective upper-triangular generators)
$(\sigma^+,\sigma_{\mathrm{W}},\sigma^-)$.

A similar argument shows that the operator
\begin{equation}
  \Psi'_\beta(z) =
  \gammaW(z)\Psi(z)e^{-\half\beta\int\limits^{\;\;z}\! J^0} 
  =  e^{f + (1 - \half\beta)\phi - 
    \frac{(k + 2) \beta - 2}{2 \sqrt{2(k + 2)}}\varphi}(z)
\end{equation}
is the $(2,2,1)$ representation of (the quantum $\SL2$ groups with the
respective upper-triangular generators)
$(\sigma^+,\sigma_{\mathrm{W}},\sigma^-)$.

\subsection{Symmetric bosonization and the triplet operator}
\label{sec:sym-bos} The above results do not depend on a particular
bosonization of the $\betaW\gammaW$ system; from now on, we use the
alternative bosonization where
\begin{equation}\label{second-bos}
  \betaW=-\d f\,e^{-f-\phi},\qquad  \gammaW=e^{f+\phi}
\end{equation}
(for the $\smash{\hSL2}$ algebra, this gives the so-called
\textit{symmetric} three-boson realization).  The point is
that \textit{all the three operators} $(S^+,S^-,S_{\mathrm{W}})$ then
take the form $(\mathrm{current})\cdot(\mathrm{exponential})$,
\begin{align}\label{Splus}
  S^+ &= \d f\,e^{(\beta-1)\phi}e^{\beta\sqrt{\frac{k+2}{2}}\varphi}
  e^{-f},\\
  S_{\mathrm{W}} &=\d f\,
  e^{-\phi} e^{-\sqrt{\frac{2}{k + 2}}\varphi}e^{-f},\\
  S^- &= \Bigl((k+1)\d f + (k+2)\d\phi +
  \sqrt{2(k+2)}\d\varphi\Bigr)e^{(\beta'+1)\phi}
  e^{\beta'\sqrt{\frac{k+2}{2}}\varphi}e^{f}.\label{Sminus}
\end{align}
In addition, it is straightforward to show that the operator
\begin{equation}\label{eq:triplet}
  \Upsilon_\beta=\bigl(4(\beta - 1)\d\phi
  + 2\sqrt{2(k + 2)}\beta\d\varphi
  + (k\beta^2 + 4\beta - 4) \d f\bigr)
  e^{\phi + \sqrt{\frac{2}{k + 2}}\varphi + f}
\end{equation}
is the $(1,3,1)$ representation of the quantum $\SL2$ groups with the
respective upper-triangular generators
$(\sigma^+,\sigma_{\mathrm{W}},\sigma^-)$.

We note that in this bosonization, the energy-momentum tensor with
central charge~\eqref{c-perp} in the commutant of the screenings is
explicitly given by
\begin{equation}\label{T-perp}
  T_\perp = \thalf\d f\d f - \thalf \d\phi\d\phi +
  \thalf\d\varphi\d\varphi - \thalf\d^2 f
  + \tfrac{k \beta^2 - (k - 4) \beta - 2}{2k\beta + 4}\d^2\phi +
  \tfrac{k(k + 2)\beta^2 + 2(k + 4)\beta - 4}{
    \sqrt{2(k + 2)}(2k\beta + 4)} \d^2\varphi.
\end{equation}

Using the symmetric bosonization, we next identify the above
screenings with the bosonic screenings constructed in
Sec.~\ref{sec:varphi123}, which also are of the form
$(\mathrm{current})\cdot(\mathrm{exponential})$.

\subsubsection{Mapping onto $(\varphi_1,\varphi_2,\varphi_3)$}
\label{sec:mapping1} Writing $S^+=(\dots)e^{\Vec r^+\cdot\Vec\Phi}$,
$S_{\mathrm{W}}=(\dots)e^{\Vec r_{\mathrm{W}}\cdot\Vec\Phi}$, and
$S^-=(\dots)e^{\Vec r^-\cdot\Vec\Phi}$, where the dots denote currents
and $\Vec\Phi=(\phi, \varphi,f)$, we note that the three 3-dimensional
vectors in the exponents are pairwise orthogonal and satisfy
\begin{equation}
  \Vec r^+\cdot \Vec r^+ = \thalf\beta(k\beta + 4),
  \qquad
  \Vec r_{\mathrm{W}}\cdot\Vec r_{\mathrm{W}} = \tfrac{2}{k+2},
  \qquad
  \Vec r^+\cdot \Vec r^+ = \thalf\beta'(k\beta' - 4) =
  -\tfrac{2\beta(k\beta + 4)}{(k\beta + 2)^2}.
\end{equation}

The identification between the
screenings in~\eqref{Splus}--\eqref{Sminus} and those in
Sec.~\ref{sec:R-scr} can be chosen such that
\begin{equation}
  \Vec r^+\cdot \Vec r^+ = \Vec r_{13}\cdot \Vec r_{13},\qquad
  \Vec r^-\cdot \Vec r^- = \Vec r_{12}\cdot \Vec r_{12}.
\end{equation}
These two equations imply
\begin{align}
  k &= -k_1-k_2-4,\\
  \beta &= \frac{2}{k_1 + 2 + \sqrt{-(k_1 + 2)(k_2 + 2)}}.
  \label{beta-1}
\end{align}
With these $k$ and $\beta$, \textit{the central charge in
  Eq.~\eqref{c-perp} becomes the one in Eq.~\eqref{the-c}}.  Further,
equating the respective exponentials in
\begin{equation}
  R_{13}= S^+,\qquad R_{23} = S_{\mathrm{W}}, \qquad R_{12} = S^-,
\end{equation}
we obtain a system of three equations that is solved by
\begin{align*}
  \d\phi = {}& -\tfrac{2 \sqrt{-(k_1 + 2) (k_2 + 2)} + (k_1 + 2) (k_1
    + k_2 + 2)}{k_1 + k_2 + 4}\d\varphi_1 + i\tfrac{2(k_2 + 1)
    \sqrt{-(k_1 + 2) (k_2 + 2)} + (k_1 + 3)(k_2 + 2)(k_1 + k_2 +
    2)}{\sqrt{k_2}
    \sqrt{k_2 + 2} (k_1 + k_2 + 4)}\d\varphi_2 \\*
  &- \tfrac{\sqrt{2(k_1 + k_2 + 2)} (\sqrt{-(k_1 + 2) (k_2 + 2)} +
    (k_1 + 2) (k_1 + k_2 + 3))}{
    \sqrt{k_1 + 2}\sqrt{k_2} (k_1 + k_2 + 4)}\d\varphi_3,\notag\\
  \d\varphi = {}& -i\tfrac{\sqrt{2(k_1 + k_2 + 2)} (k_1 + 2 +
    \sqrt{-(k_1 + 2) (k_2 + 2)})}{ k_1 - k_2 + 2 \sqrt{-(k_1 + 2) (k_2
      + 2)}}\d\varphi_1 - \tfrac{\sqrt{2(k_1 + k_2 + 2)} ((k_1 +
    1)(k_2 + 2) + \sqrt{-(k_1 + 2) (k_2 + 2)} (k_2 + 3))}{ \sqrt{k_2}
    \sqrt{k_2 + 2} (k_1 - k_2 + 2
    \sqrt{-(k_1 + 2) (k_2 + 2)})}\d\varphi_2\\
  &+ i\tfrac{\sqrt{-(k_1 + 2)(k_2 + 2)} (k_1 + k_2 + 2) - 2(k_1 + 2)
    (k_1 + k_2 + 3)}{
    \sqrt{k_1 + 2}\sqrt{k_2}(k_1 + k_2 + 4)}\d\varphi_3,\notag\\
  \d f = {}& (k_1 + 1)\d\varphi_1 - i(k_1 + 2)\sqrt{\tfrac{k_2 +
      2}{k_2}}\d\varphi_2 + \tfrac{\sqrt{2(k_1 + 2)} \sqrt{k_1 + k_2 +
      2}}{ \sqrt{k_2}}\d\varphi_3
\end{align*}

In terms of the three currents introduced in Sec.~\ref{sec:varphi123},
the integrands of the screenings become
\begin{align}
  R_{13} = S^+&= \Vec a_3\cdot\d\Vec\varphi\,e^{\Vec
    r_{13}\cdot\Vec\varphi},\\ 
  R_{23} = S_{\mathrm{W}}&= -\Vec a_3\cdot\d\Vec\varphi\,
  e^{\Vec r_{23}\cdot\Vec\varphi},\\
  R_{12} = S^-&= \Vec a_2\cdot\d\Vec\varphi\,
  e^{\Vec r_{12}\cdot\Vec\varphi}
\end{align}
and the ``deformed'' vertex operators are mapped into
\begin{align}
  \Psi_\beta&=
  e^{-\half(\Vec r_{12} + \Vec r_{23})\cdot\Vec\varphi},\\
  \Psi'_\beta&=
  e^{-\half(\Vec r_{13} + \Vec r_{23})\cdot\Vec\varphi}.
\end{align}
Their dimensions evaluated with respect to the energy-momentum tensor
in Eq.~\eqref{T-perp} are expressed in terms of $k_1$ and $k_2$ as
\begin{equation}\label{dimensions}
  \dim \Psi_\beta =
  \tfrac{3k_1}{4 (k_2 + 2) (k_1 + k_2 + 2)},
  \qquad
  \dim \Psi'_\beta = \tfrac{3 k_2}{4 (k_1 + 2) (k_1 + k_2 + 2)}.
\end{equation}

The $\Upsilon_\beta$ operator in Eq.~\eqref{eq:triplet} becomes, up to
normalization,
\begin{equation}\label{eq:triplet-3}
  \Upsilon_\beta=\Vec a_1\cdot\d\Vec\varphi\,
  e^{-\Vec r_{23}\cdot\Vec\varphi}.
\end{equation}
We note that its property to be the $(1,1,3)$ representation with
respect to $(\rho_{12},\rho_{13},\rho_{23})$ is now readily checked
using~\eqref{useful}.

\subsubsection{Identification with $\WW\!\D$} We now identify the
$W_*$ algebra in the commutant of the
$(\sigma^+,\sigma_{\mathrm{W}},\sigma^-)$ screenings with $\WW\!\D$.
In the symmetric bosonization of $\hSL2$, there is one bosonic
(Wakimoto) and two fermionic screenings.  The $W_*$ algebra is
therefore selected from the free-field space by five screening
operators.  This set of screenings is redundant: as we have seen,
$W_*$ is in fact the commutant of $\sigma^+$, $\sigma_{\mathrm{W}}$,
and~$\sigma^-$.  On the other hand, it is the commutant of the two
fermionic screenings and $\sigma^+$ (equivalently, of the two
fermionic screenings and $\sigma^-$).  This follows from viewing $W_*$
as the intersection of two algebras of the form
$\WW\SSL21\tensor(\mathrm{Heisenberg})$.  These three screenings,
however, have been identified with the respective screenings in
Sec.~\ref{sec:varphi123}.  It only remains to verify the
identification for the fermionic screenings in the symmetric
bosonization of~$\hSL2$,
\begin{equation*}
  e^{(k + 2)\phi + \sqrt{2(k + 2)}\varphi
    + (k + 1)f}\quad\text{and}\quad e^{f}.
\end{equation*}
Under the mapping in Sec.~\ref{sec:mapping1}, these are indeed mapped
into $\sigma_2$ and $\sigma_3$ in
Eqs.~\eqref{sigma-2-final}--\eqref{sigma-3-final}.  We thus conclude
that the $W$ algebra in the commutant of
$(\sigma^+,\sigma_{\mathrm{W}},\sigma^-)$ coincides with~$\WW\!\D$.

\bigskip

Thus, we have constructed two vertex operators for $\WW\!\D$ such that
each vertex is a singlet with respect to one of the three $\uU_q\SL2$
quantum groups and is (the highest-weight vector in) a doublet with
respect to each of the other two.  These operators are used in
reconstructing the spin-$\half$ vertex operators of $\hSL2_{k_1}$ and
$\hSL2_{k_2}$.  In reconstructing the currents, we also use the
operator in Eq.~\eqref{eq:triplet-3}.  We change the notation for
these operators as
\begin{equation}\label{the-three}
  \Psi_{212} =
  e^{-\half(\Vec r_{12} + \Vec r_{23})\cdot\Vec\varphi},\qquad
  \Psi_{122} =
  e^{-\half(\Vec r_{13} + \Vec r_{23})\cdot\Vec\varphi},\qquad
  \Upsilon_{113}=\Vec a_1\cdot\d\Vec\varphi\,
  e^{-\Vec r_{23}\cdot\Vec\varphi},
\end{equation}
where the subscripts refer to representations of the $\uU_q\SL2$
quantum groups with the respective upper-triangular generators
$\rho_{12}$, $\rho_{13}$, and $\rho_{23}$.

\section{The vertex operator reconstruction} \label{sec:reconstruct}
In this section, we reconstruct $\hSL2_{k_1}$ and $\hSL2_{k_2}$
currents and vertex operators using $\WW\!\D$ vertex operators.  This
will imply that $\WW\!\D$ is the $W$ algebra of the
$\hSL2_{k_1}\oplus\hSL2_{k_2}/\hSL2_{k_1 + k_2}$ coset.  Because we
use the language of vertex-operator extensions, we start with
reviewing several relevant points in Sec.~\ref{sec:qgs-and-vertices}.
In Sec.~\ref{sec:coset-recall}, we recall defining properties of the
$\hSL2\oplus\hSL2/\hSL2$ coset theory (originating
in~\cite{[GK],[KS]}, see also~\cite{[HR]}) and then show that these
can be recovered starting from the $\WW\!\D$ algebra.  In
Sec.~\ref{sec:reconstruct-currents}, we reconstruct the $\hSL2_{k_1}$
and $\hSL2_{k_2}$ currents, and in Sec.~\ref{subsec:reconstruct},
vertex operators.

\subsection{Quantum groups and vertex-operator extensions}
\label{sec:qgs-and-vertices} Vertex-operator extensions $\cA\to\cB$,
e.g., the one that we consider in~\eqref{the-extension} in what
follows, are constructed by adding \textit{local} fields to the
algebra $\cA$; such fields are selected as those vertex operators
of~$\cA$ that have trivial monodromies with respect to each other.
Because the monodromy properties are encoded in quantum group
representations, the procedure involves taking quantum-group singlets
that are simply transposed under the action of the corresponding
$R$-matrix.  For the particular type of vertex-operator extensions
that we use, the $\cA$ algebra is taken to be a product
$\cA=\cA^{(1)}\tensor\cA^{(2)}$, where each algebra $\cA^{(i)}$ is
considered together with an algebra of its vertex operators; from the
$\cA^{(1)}$ and $\cA^{(2)}$ vertex operators, one then constructs all
possible \textit{local} fields (i.e., fields with trivial monodromies
with respect to each other).  A commonly used recipe to construct
local fields with the help of $R$ matrices amounts to finding vertex
operators $V^{(i)}_\alpha$ of the respective algebra $\cA^{(i)}$ and
the elements
\begin{equation}
  w_{\alpha\alpha'}\in V^{(1)}_\alpha\tensor V^{(2)}_{\alpha'}
\end{equation}
(where we somewhat abuse the notation by identifying vertex operators
with the corresponding quantum-group representation spaces) such that
the monodromy properties
\begin{equation}
  R^{(i)}_{\alpha\beta}: V^{(i)}_\alpha\tensor V^{(i)}_\beta \to
  V^{(i)}_\beta\tensor V^{(i)}_\alpha
\end{equation}
imply that the mapping
\begin{align}
  P_{23}\circ R^{(1)}_{\alpha\beta}\tensor
  R^{(2)}_{\alpha'\beta'}\circ P_{23}:
  V^{(1)}_\alpha\tensor V^{(2)}_{\alpha'} \tensor
  V^{(1)}_\beta\tensor V^{(2)}_{\beta'}&\to
  V^{(1)}_\beta\tensor V^{(2)}_{\beta'} \tensor
  V^{(1)}_\alpha\tensor V^{(2)}_{\alpha'},\\
\intertext{reduces to transpositions of the chosen elements,}
\label{no-monodromy}
  P_{23}\circ R^{(1)}_{\alpha\beta}\tensor
  R^{(2)}_{\alpha'\beta'}\circ P_{23}:
  w_{\alpha\alpha'}\tensor w_{\beta\beta'}&\mapsto
  w_{\beta\beta'}\tensor w_{\alpha\alpha'}.
\end{align}

All such operators $w_{\alpha\alpha'}$ then make up a local algebra,
which is a vertex-operator extension $\cB$ of
$\cA=\cA^{(1)}\tensor\cA^{(2)}$.  Next, (some of) the \textit{vertex
  operators} of~$\cB$ can be constructed similarly, by combining
$\cA^{(1)}$ and $\cA^{(2)}$ vertex operators into operators that have
some prescribed monodromy properties instead of the trivial monodromy
in~\eqref{no-monodromy} (and are local with \hbox{respect to the
  elements of}~$\cB$).

In what follows, we let~$\oC^n_q$ denote the $n$-dimensional module
over the quantum group~$\uU_q\SL2$ (we use these with $n=2$ and~$3$);
abusing the notation, we also write $\oC^2_{\kappa}$ whenever
$q=e^{\pi i\kappa}$.  The relevant value of the~$q$ parameter for each
quantum group is read off from the corresponding screening operator.
We recall from Sec.~\ref{sec:mapping1} the vertex operators
$\Psi_{212}$ and $\Psi_{122}$ that are the $(2,1,2)$ and $(1,2,2)$
representations of the $\uU_q\SL2$ quantum groups with the
upper-triangular generators $\rho_{12}$, $\rho_{13}$, and~$\rho_{23}$.
It follows from the formulae of Sec.~\ref{sec:varphi123} that the
respective quantum group parameters are $q_j=e^{\pi i\kappa_j}$ with
\begin{equation}\label{the-kappa}
  \kappa_1 = \frac{2}{k_2 + 2},\qquad
  \kappa_2 = \frac{2}{k_1 + 2},\qquad
  \kappa_3 = \frac{-2}{k_1 + k_2 + 2}.
\end{equation}
Abusing the terminology, we often extend the notations $\Psi_{212}$
and $\Psi_{122}$ to the respective 4-dimension\-al quantum-group
representations, with the specific operators in the right-hand sides
of~\eqref{the-three} being the highest-weight vectors in these
representations.  The $(2,1,2)$ representation content of $\Psi_{212}$
can therefore be expressed as
\begin{align}
  \Psi_{212}(z)&\in(\oC^2_{\frac{2}{k_2 + 2}}
  \tensor\oC^2_{-\frac{2}{k_1 + k_2 + 2}})(z)\\[-6pt]
  \intertext{and likewise,}
  \Psi_{122}(z)&\in(\oC^2_{\frac{2}{k_1 + 2}}
  \tensor\oC^2_{-\frac{2}{k_1 + k_2 + 2}})(z).
\end{align}
Similarly, we also write
\begin{equation}
  \Upsilon_{113}(z)\in\oC^3_{-\frac{2}{k_1 + k_2 + 2}}(z).
\end{equation}
Each of the above operators is a singlet with respect to the $\uU_q\D$
quantum group.

The monodromy properties of these $\WW\!D_{2|1}(k_1,k_2)$ vertex
operators are therefore governed by the lower-dimensional $\uU_q\SL2$
representations.  For convenience, we give our quantum group
conventions and summarize the construction of elements with trivial
monodromies, see~\cite{[Kass]}.  The $\SL2$ quantum group relations
are
\begin{align}
  K\,K^{-1} &= K^{-1}\,K =1,\notag\\
  K\,E\,K^{-1} &= q^2\,E,\qquad K\,F\,K^{-1} = q^{-2}\,F,\qquad [E, F]
  = \frac{K - K^{-1}}{q - q^{-1}}.\notag
\end{align}
with the comultiplication given by
\begin{gather}
  \Delta(E)=1\tensor E + E\tensor K,\qquad
  \Delta(F)=K^{-1}\tensor F + F\tensor 1,\\
  \Delta(K)=K\tensor K,\qquad \Delta(K^{-1})=K^{-1}\tensor K^{-1}.
\end{gather}
%
For a positive integer $n$, let $\qmV_{n}$ be the~$\uU_q\SL2$ module
with the highest-weight vector $v_0$ such that $E v_0=0$, $K v_0=
q^{n} v_0$, and $F v_{i-1}=[i] v_i$ (with the standard notation
$[i]=\frac{q^i - q^{-i}}{q-q^{-1}}$).  It follows that $E v_i=
[n-i+1] v_{i-1}$ and $K v_{i} = q^{n-2i} v_i$.

The idea of constructing elements with trivial monodromies is to
combine representations of $\uU_q\SL2$ and $\uU_{q^{-1}}\SL2$ quantum
groups.  As the basic example, we consider the~$\qmV_{1}$ module.
Let $\qmV'_{1}$ be a similar module over $\uU_{q^{-1}}\SL2$, with the
basis $v'_0$ and $v'_1$. \textit{It is also a module over
  $\uU_q\SL2$}, with the~$\uU_q\SL2$ action given by
\begin{alignat}{3}\label{q-inverse1}
  E\,v'_0={}&v'_1,&\qquad K\,v'_0={}&q^{-1}\,v'_0,&\qquad
  F\,v'_0={}&0,\\
  E\,v'_1={}&0,& K\,v'_1={}&q\,v'_1,&
  F\,v'_1={}&v'_0.\label{q-inverse2}
\end{alignat}
The tensor product of~$\uU_q\SL2$ modules $\qmV'_{1}\tensor \qmV_{1}$
is decomposed as
\begin{equation}
  \qmV'_{1}\tensor \qmV_{1} =
  \oC\oplus \qmV_{2},
\end{equation}
where $\qmV_{2}$ is generated from $v'_1\tensor v_0$, and
$\qmV_{0}=\oC$ from $w_2=v'_0\tensor v_0 - q v'_1\tensor v_1$.  
This gives an element of the ``$w_{\alpha\alpha'}$-type'' in
Eq.~\eqref{no-monodromy}. 
Similar formulae can be easily written for the invariant element
$w_3\in\qmV'_{2}\tensor\qmV_{2}$ and
$w_{n+1}\in\qmV'_{n}\tensor\qmV_{n}$.

\subsection{The
  $\smash{\protect\hSL2\oplus\protect\hSL2/\protect\hSL2}$ coset}
\label{sec:coset-recall} This coset conformal field theory can be
defined as the relative semi-infinite cohomology
$H^{\infty/2}(\hSL2_{-4},\SL2)$ of the complex
\begin{equation}\label{symmetric}
  \hSL2_{k_1}\oplus\hSL2_{k_2}\oplus\hSL2_{k_3}\oplus
  \text{ghosts}
  \quad\text{with}\quad k_1 + k_2 + k_3 = -4,
\end{equation}
where the ghosts are given by three free-fermion (``$BC$'') systems
and the differential is constructed in the standard way for the
diagonally embedded level-$(-4)$ $\hSL2$ algebra~\cite{[F],[FGZ]} (see
also~\cite{[H]} and references therein) starting with the differential
\begin{equation}
  d = \oint(J^+ C_+ + J^- C_- + J^0 C_0 - 2 B^0 C_+ C_- -
  B^- C_- C_0 + B^+ C_+ C_0)
\end{equation}
that computes the absolute cohomology.  The differential can be
defined to act on the tensor product of the ghost modules and the
vacuum representations of the three $\hSL2$ algebras.  For generic
$k_1$ and $k_2$, the cohomology of this complex is concentrated in the
ghost number zero, and the vertex operator algebra in the cohomology
is the conformal field theory associated with the coset.  This gives
\textit{local fields} (vacuum descendants)---the elements of
$H^{\infty/2}(\hSL2_{-4},\SL2;
\Vac_{(k_1)}\tensor\Vac_{(k_2)}\tensor\Vac_{(k_3)})$.

Taking cohomology elements with more general coefficients gives vertex
operators of this coset theory rather than just the local fields.  An
obvious class $\cC_0$ of the $\hSL2_{k_1}\oplus\hSL2_{k_2}/\hSL2_{k_1
  + k_2}$ vertex operators are constructed as follows.  One starts
with vertex operators for each $\hSL2$, which are represented by
$\oC^m(z)\tensor\oC_q^m$, where the first~$\oC^m$ specifies the
horizontal $\SL2$ subalgebra representation and $\oC_q^m$ the
$\uU_q\SL2$ quantum group representation.  The reduction to the coset
then consists in selecting those operators that commute with the above
differential, which amounts to taking $\SL2$ invariants in
$\oC^m\tensor\oC^n\tensor\oC^\ell$.  This means selecting the triples
$(m,n,\ell)$ such that the tensor product
$\oC^m\tensor\oC^n\tensor\oC^\ell$ of $\SL2$ representations contains
an invariant with respect to the diagonal~$\SL2$ algebra.  This
therefore gives the $\hSL2_{k_1}\oplus\hSL2_{k_2}/\hSL2_{k_1 + k_2}$
vertex operators that are in a \hbox{$1:1$} correspondence with the
space of invariants $[\oC^m\tensor\oC^n\tensor\oC^\ell]^{\SL2}$.

Reformulating the above construction in terms of the semi-infinite
cohomology immediately suggests a more general construction that gives
a larger class of coset vertex operators involving descendants.
Descendants of the $(\oC^m\tensor\oC^n\tensor\oC^\ell)(z)$ vertex
operators span the tensor product of the corresponding Weyl modules,
\begin{equation*}
  (\cM_m\tensor\cM_n\tensor\cM_\ell)(z),
\end{equation*}
where $\cM_i$ is the Weyl module with the $(i+1)$-dimensional
top-level $\SL2$ representation and the three tensor factors are
representations of $\hSL2_{k_1}$, $\hSL2_{k_2}$, and $\hSL2_{k_3}$,
respectively.  The reduction then consists in calculating the
cohomology spaces
\begin{equation}\label{coh-vertex}
  H^{\infty/2}(\hSL2_{-4},\SL2;\cM_m\tensor\cM_n\tensor\cM_\ell)(z)
\end{equation}
for each triple of positive integers $(m,n,\ell)$, which again amounts
to a similar problem of finding $\SL2$ invariants (these are more
numerous once further Weyl-module levels are involved beyond the top
ones).

The space $\bigoplus_{m,n,\ell}H^{\infty/2}(\hSL2_{-4},\SL2;
\cM_m\tensor\cM_n\tensor\cM_\ell)(z)$ is closed under operator
products and represents a subalgebra of the algebra of the
$\hSL2_{k_1}\oplus\hSL2_{k_2}/\hSL2_{k_1 + k_2}$ vertex operators.
This space, moreover, carries a representation of local fields of the
coset conformal conformal field theory because the algebra
$H^{\infty/2}(\hSL2_{-4},\SL2;\Vac\tensor\Vac\tensor\Vac)$ acts on
$\bigoplus_{m,n,\ell} H^{\infty/2}(\hSL2_{-4},\SL2;
\cM_m\tensor\cM_n\tensor\cM_\ell)$.\footnote{From the identification
  of the coset with the $W$ algebra that we obtain below, it follows
  that the above construction describes those coset vertex operators
  that are singlets with respect to the $\uU_q\D$ quantum group; only
  these follow in an obvious way from the definition of the coset,
  whereas the construction of vertex operators carrying nontrivial
  $\uU_q\D$ representations is left for the future work.}

The definition of the coset as $W=H^{\infty/2}(\hSL2_{-4},\SL2;
\Vac_{(k_1)}\tensor\Vac_{(k_2)}\tensor\Vac_{(k_3)})$, with its vertex
operators in
$H^{\infty/2}(\hSL2,\SL2;\cM_m\tensor\cM_n\tensor\cM_\ell)$, leads to
another property characterizing the coset theory: the product of the
coset with the $\hSL2$ algebra of the level $-k_i-4$ that is ``dual''
to~$k_i$ for any $i=1,2,3$ admits a vertex-operator extension to the
product of the other two $\hSL2_{k_i}$ algebras:
\begin{equation}\label{with-dual-level}
  W\tensor\uU\hSL2_{-k_1-4}\xrightarrow{\text{v.o.e.}}
  \uU\hSL2_{k_2}\tensor\uU\hSL2_{k_3}
\end{equation}
(where $\uU$ denotes vacuum representations).  This vertex-operator
extension is achieved by combining the coset vertex operators
described above and the standard $\hSL2_{-4}$ vertex operators.  The
relevant mapping is in fact given by the isomorphism
\begin{equation}
  \bigoplus_{n\geq0}H^{\infty/2}(\hSL2,\SL2;
  \cM_n\tensor\Vac_{(k_2)}\tensor\Vac_{(k_3)})
  \tensor\cM_n\xrightarrow{\sim}
  \Vac_{(k_2)}\tensor\Vac_{(k_3)}
\end{equation}
(with the two Weyl modules being those over the $\hSL2_{k_1}$ and
$\hSL2_{-k_1-4}$ algebras), which involves the contraction ``over the
quantum group index'' induced by taking the monodromy-free element in
each term.

Property~\eqref{with-dual-level} characterizes $W$ as the coset,
because $\hSL2_{-k_1-4}$ is then \textit{diagonally} embedded in
$\hSL2_{k_2}\oplus\hSL2_{k_3}$ (we recall that $k_1+k_2+k_3=-4$), and
therefore,
\begin{equation}
  W\tensor\uU\hSL2_{-k_1-4}\tensor
  \uU\hSL2_{k_1} 
  \longrightarrow
  \uU\hSL2_{k_1}\tensor\uU\hSL2_{k_2}\tensor\uU\hSL2_{k_3},
\end{equation}
where on the right-hand side, the $\hSL2_{-4}$ algebra is embedded
diagonally.  In accordance with the definition, therefore, the relative
semi-infinite cohomology of the right-hand side reproduces the coset,
while on the left-hand side, $H^{\infty/2}(\hSL2_{-4},\SL2)$ evaluates
as $\oC$ on $\uU\hSL2_{-k_1-4}\times \uU\hSL2_{k_1}$ and thus gives
$W$.

It is the property in Eq.~\eqref{with-dual-level} that we show
for~$\WW\!D_{2|1}(k_1,k_2)$ instead of~$W$.  Rewriting this with $k_1$
replaced by $k_3=-k_1-k_2-4$, we construct in
Sec.~\ref{sec:reconstruct-currents} the vertex-operator extension
\begin{equation}\label{the-extension}
  \WW\!D_{2|1}(k_1,k_2)\tensor\uU\SL2_{k_1+k_2}
  \xrightarrow{\text{v.o.e.}}\uU\hSL2_{k_1}\tensor\uU\hSL2_{k_2}.
\end{equation}
As we have seen, this implies a homomorphism
\begin{equation}
  \WW\!D_{2|1}(k_1,k_2)\longrightarrow
  H^{\infty/2}(\hSL2_{-4},\SL2;
  \Vac_{(k_1)}\tensor\Vac_{(k_2)}\tensor\Vac_{(k_3)}),
\end{equation}
which is in fact an isomorphism.  Thus, constructing vertex-operator
extension~\eqref{the-extension} will show that $\WW\!D_{2|1}(k_1,k_2)$
is indeed the coset $\hSL2_{k_1}\oplus\hSL2_{k_2}/\hSL2_{k_1 + k_2}$.

\subsection{Reconstructing $\smash{\hSL2_{k_1}\oplus\hSL2_{k_2}}$
  currents} \label{sec:reconstruct-currents} We now construct the
vertex-operator extension~\eqref{the-extension} by combining
$\WW\!D_{2|1}(k_1,k_2)$ vertex operators constructed in
Sec.~\ref{sec:constr-coset-deform} with $\hSL2_{k_1+k_2}$ vertex
operators.

The operator in Eq.~\eqref{eq:triplet-3} that is a 3-dimensional
representation of the $\uU_{q^{-1}}\SL2$ quantum group with
$q=e^{\frac{2\pi i}{k_1+k_2+2}}$ can be contracted with the spin-$1$
vertex operator $\Phi_1(k_1+k_2)$ for $\hSL2_{k_1+k_2}$.  The latter
is a 3-dimensional representation of $\uU_{q}\SL2$, which we express
by writing $\Phi_1(k_1+k_2)(z)=\oC^3(z)\tensor\oC^3_q$ (with the first
$\oC^3$ factor representing the triplet with respect to the horizontal
$\SL2$ subalgebra).  This contraction ``with respect to the
quantum-group indices'' amounts to taking a quantum-group singlet
$w_3\in\oC^3_q\tensor\oC^3_{q^{-1}}$.  The result is then a local
field in the sense of~\eqref{no-monodromy}.  We thus have
\begin{equation}\label{3-contraction}
  \underbrace{\oC^3(z)\tensor\oC^3_{\frac{2}{k_1 + k_2 + 2}}}_{
    \Phi_1(k_1+k_2)(z)}{}\tensor{}
  \underbrace{\oC^3_{\frac{-2}{k_1 + k_2 + 2}}(z)}_{\Upsilon_{113}(z)}
  \ni \oC^3(z)\tensor w_3.
\end{equation}
The right-hand side is therefore a 3-dimensional representation of the
$\SL2$ subalgebra of $\hSL2_{k_1+k_2}$.  In this 3-dimensional space,
we choose the basis $J_*^+(z)$, $J_*^0(z)$, and $J_*^-(z)$ such that
$J_*^+(z)$ corresponds to the highest-weight vector.  We let
$J^{\pm,0}_{\mathrm{diag}}(z)$ denote the~$\hSL2_{k_1+k_2}$ currents.

The space on the right-hand side of~\eqref{3-contraction} is in fact
the evaluation representation of~$\hSL2_{k_1+k_2}$.  Before the
contraction, a number of cancellations of poles occur in the nine
operator products in $\oC^3_{\frac{-2}{k_1 + k_2 + 2}}(z)\cdot
\oC^3_{\frac{-2}{k_1 + k_2 + 2}}(w)$ between the different components
of $\Upsilon_{113}$.  For the highest-weight elements in the
quantum-group representations explicitly written in~\eqref{the-three},
we have
\begin{equation}
  \Vec a_1\cdot\d\Vec\varphi\,
  e^{-\Vec r_{23}\cdot\Vec\varphi}(z)\,
  \Vec a_1\cdot\d\Vec\varphi\,
  e^{-\Vec r_{23}\cdot\Vec\varphi}(w)=
  (z-w)^{-\frac{2}{k_1 + k_2 + 2}}
  \Bigl(\frac{\Vec a_1\cdot\Vec a_1 - (\Vec a_1\cdot\Vec r_{23})^2}{
    (z-w)^2} +\dots \Bigr)\dots,
\end{equation}
which shows that the leading pole vanishes.  After taking the
quantum-group contraction, this implies that the currents
\begin{align}
  J^{\pm,0}_1 &= \tfrac{1}{k_1+k_2}(k_1 J^{\pm,0}_{\mathrm{diag}} +
  J^{\pm,0}_*),\\
  J^{\pm,0}_2 &= \tfrac{1}{k_1+k_2}(k_2 J^{\pm,0}_{\mathrm{diag}} -
  J^{\pm,0}_*)
\end{align}
satisfy the respective $\hSL2_{k_i}$ algebras.\footnote{And therefore,
  $J_*^{\pm,0}$ constructed above can be identified with the
  representation of $\hSL2_{k_1+k_2}$ spanned by the $k_2
  J^{\pm,0}_1(z) - k_1 J^{\pm,0}_2(z)$ currents in
  $\hSL2_{k_1}\oplus\hSL2_{k_2}$.}  This shows~\eqref{the-extension}.

\subsection{Reconstructing $\smash{\hSL2_{k_i}}$ vertex operators}
\label{subsec:reconstruct} Next,
$\hSL2_{k_1}\oplus\hSL2_{k_2}$ vertex operators also follow from a
vertex-operator extension.  Let
$\Phi_{\half}(k)(z)=\oC^2(z)\tensor\oC_{\frac{2}{k+2}}^2$ be the
vertex operators for the spin-$\half$ representation of $\hSL2_{k}$.
Here,~$\oC^2_{\frac{2}{k+2}}$ is the two-dimensional representation of
$\uU_q\SL2$ with the quantum group parameter~$q=e^{2 \pi i/(k + 2)}$
and the first $\oC^2$ factor represents the $\SL2$ doublet.

We now combine this vertex with vertex operators for $\WW
D_{2|1}(k_1,k_2)$ that transform under the two-dimensional
representation of~$\uU_{q^{-1}}\SL2$.  In the tensor product, there
exists an $\uU_q\SL2$ invariant element (in the notation of
Sec.~\ref{sec:qgs-and-vertices})
\begin{equation}
  \oC^2_{q}\tensor\oC^2_{q^{-1}} \ni
  w_2=v'_0\tensor v_0 - q v'_1\tensor v_1.
\end{equation}
Applying this to $\Phi_{\half}(k_1+k_2)$ and $\Psi_{212}$ as
\begin{align}\label{reconstruct1}
  \underbrace{\oC^2(z)\tensor
    \oC_{\frac{2}{k_1+k_2+2}}^2}_{\Phi_{\half}(k_1+k_2)(z)}
  {}\tensor{}\underbrace{
    (\oC^2_{\frac{2}{k_2+2}}\tensor\oC^2_{\frac{-2}{k_1+k_2+2}})(z)
    }_{\Psi_{212}(z)} &{}
  \ni
  \oC^2(z)\tensor\oC_{\frac{2}{k_2+2}}^2\tensor w_2
  \\[-11pt]
  \intertext{and similarly to $\Phi_{\half}(k_1+k_2)$ and
    $\Psi_{122}$,}
  \underbrace{\oC^2(z)\tensor
    \oC_{\frac{2}{k_1+k_2+2}}^2}_{\Phi_{\half}(k_1+k_2)(z)}
  {}\tensor{}\underbrace{
    (\oC^2_{\frac{2}{k_1+2}}\tensor\oC^2_{\frac{-2}{k_1+k_2+2}})(z)
    }_{\Psi_{122}(z)}&
  \ni
  \oC^2(z)\tensor\oC_{\frac{2}{k_1+2}}^2\tensor w_2,
  \label{reconstruct2}
\end{align}
we obtain the respective $\oC^2(z)\tensor\oC^2_q$ structures that can
be identified with the $\hSL2_{k_2}$ and $\hSL2_{k_1}$ spin-$\half$
vertex operators $\Phi_{\half}(k_2)(z)$ and $\Phi_{\half}(k_1)(z)$.
In fact, monodromy properties alone do not guarantee that the
operators constructed are necessarily primary.  Showing that they are
primary involves examining the relevant operator products.
Pole cancellations occur in $\oC^3_{-\frac{2}{k_1 + k_2 + 2}}(z)\cdot
(\oC^2_{\frac{2}{k_2 + 2}}\tensor \oC^2_{-\frac{2}{k_1 + k_2 +
    2}})(w)$ in acting with $\Upsilon_{113}(z)$ on $\Psi_{212}(w)$,
and similarly for~$\Psi_{122}$.  For the highest-weight vectors of the
respective quantum-group multiplets, we have
\begin{equation}
  \Vec a_1\cdot\d\Vec\varphi\,
  e^{-\Vec r_{23}\cdot\Vec\varphi}(z)\;
  e^{-\half(\Vec r_{12} + \Vec r_{23})\cdot\Vec\varphi}(w)=
  (z-w)^{-\frac{1}{k_1 + k_2 + 2}}
  \Bigl(\frac{-\half\Vec a_1\cdot(\Vec r_{12} + \Vec r_{23})}{z-w} +
  \dots\Bigr)\dots,
\end{equation}
which shows that the leading pole vanishes in view of~\eqref{useful}.
The entire vertex operator algebra of each $\hSL2_{k_i}$ can be
obtained similarly (or simply by noting that components of the
spin-$\half$ operator generate the entire vertex operator algebra).


Thus, we have shown that the $\WW\!D_{2|1}(k_1,k_2)$ algebra defined
as the commutant of $\uU_q\D$ is the coset
$\hSL2_{k_1}\oplus\hSL2_{k_2}/\hSL2_{k_1 + k_2}$.  We next show that
the same $W$ algebra can be obtained by the Hamiltonian reduction of
the affine Lie superalgebra $\hD$.

\section{Hamiltonian reduction of $\hD$} \label{sec:Ham-red}
In this section, we construct the Hamiltonian reduction
$\smash{\hD\to\hSL2\oplus\hSL2/\hSL2}$.

\subsection{Remarks on Hamiltonian reduction} For a Lie (super)algebra
$\aG$, the general setting of the Hamiltonian reduction of
$\widehat\aG$ can be formulated as follows.  For a chosen maximal
nilpotent subalgebra $\aN\subset\aG$, one fixes an associative algebra
$\cA$ and a homomorphism $\uU\widehat\aN\to\cA$ of associative
algebras.  One then performs the reduction to
$\uU\widehat\aG\times\cA/\widehat\aN$, i.e., the reduction with
respect to the diagonal embedding
\begin{equation}\label{reduction}
  \begin{array}{c}
    \uU\widehat\aG\tensor\cA\\[2pt]
    \bigm\uparrow\\[4pt]
    \widehat\aN
  \end{array}
\end{equation}
This amounts to introducing a ``BRST'' operator (of the type used in
the semi-infinite cohomology) implementing the constraints that state
the vanishing of the diagonally embedded $\aN$-valued currents.  For
example, the Hamiltonian reduction of $\hSL2$ is reformulated in this
way with $\cA=\oC$ and the morphism given by $J^+(z)\mapsto-1$; the
embedding is therefore given by $J^+(z)\mapsto(J^+(z), -1)$.  In more
general cases, $\cA$ can be a free-field algebra.  ``Partial''
Hamiltonian reductions (with only a part of the Chevalley generators
constrained) can also be reformulated in this way by appropriately
choosing the mapping $\widehat\aN\to\cA$.

Unlike for bosonic affine Lie algebras, there is no preferred
reduction scheme for superalgebras.  For $\aG=\D$, there are several
``natural'' choices of~$\cA$ and of the homomorphism (an additional
source of potentially different reductions is due to inequivalent
choices of the maximal nilpotent subalgebra, see also
Sec.~\ref{BFB-scr}).  We consider the scheme of type~\eqref{reduction}
leading to the $W$ algebra $\WW\!\D$ defined in
Sec.~\ref{sec:commutant}.  This reduction scheme is asymmetric with
respect to the three fermionic roots.\footnote{It may also be
  interesting to study the ``symmetric'' reduction, where $\cA$ is
  generated by three free-fermion systems $b_1,c_1$, $b_2,c_2$, and
  $b_3,c_3$ (with $b_i(z)c_j(w)=\frac{\delta_{ij}}{z-w}$) and the
  mapping of the nilpotent subalgebra into $\cA$ is given by
  $\psi(\m,\p,\p)\mapsto b_1$, $\psi(\p,\m,\p)\mapsto b_2$,
  $\psi(\p,\p,\m)\mapsto b_3$, $e^{(i)}\mapsto0$ for $i=1,2,3$, and
  $\psi(\p,\p,\p)\mapsto0$. This (otherwise natural) choice of the
  reduction scheme does not lead to the $\WW\!\D$ algebra, and we do
  not expect any reduction that is symmetric with respect to the three
  fermionic roots to result in~$\WW\!\D$.}

\subsection{Hamiltonian reduction of $\hD$} We now show that
for generic values of the parameters, applying a scheme of
type~\eqref{reduction} to $\hD$ gives the product of a Heisenberg
algebra~$\cH_0$ and the $W$ algebra~$\WW\!\D$.  The Heisenberg algebra
(a free field theory) is ``trivial'' piece in that it is guaranteed by
the choice of~$\cA$, while the identification of the $\WW\!\D$ algebra
as the part of the cohomology commuting with~$\cH_0$ is the main
result of this section.

With the notation for the $\hD$ algebra introduced in
Appendix~\ref{app:D-all}, we now for brevity denote the fermionic
currents corresponding to the Chevalley generators as
\begin{equation}
  \psi_1 = \psi(\m,\p,\p),\qquad \psi_2 = \psi(\p,\m,\p),
  \qquad \psi_3 = \psi(\p,\p,\m),
  \qquad \psi_0 = \psi(\p,\p,\p).
\end{equation}
In~\eqref{reduction}, we take $\cA$ to be the algebra generated by a
free fermion system $\eta,\xi$ with the operator product
\begin{equation}
  \eta(z)\,\xi(w)=\frac{1}{z-w}
\end{equation}
with the mapping of the nilpotent subalgebra currents to $\cA$ given
by $\psi_1\mapsto -\eta$, $\psi_2\mapsto -\eta$, $\psi_3\mapsto -\xi$
and accordingly, $e^{(1)}\mapsto \tfrac{1}{2 \A_1}$,
$e^{(2)}\mapsto\tfrac{1}{2 \A_2}$, $e^{(3)}\mapsto0$, and
$\psi_0\mapsto0$.  In terms of constraints, this is
\begin{equation}\label{the-constraints}
  \psi_1(z) - \eta(z) = 0,\qquad \psi_2(z) - \eta(z) = 0,\qquad
  \psi_3(z) - \xi(z) = 0,
\end{equation}
which implies
\begin{equation}
  e^{(1)}(z) + \frac{1}{2 \A_1} = 0, \qquad
  e^{(2)}(z) + \frac{1}{2 \A_2} = 0, \qquad
  e^{(3)}(z) = 0,\qquad
  \psi_0(z) = 0.
\end{equation}

The corresponding BRST operator is given by $\cQ=\oint\cJ$ with
\begin{multline}
  \cJ = (\psi_1 - \eta) \ggamma_1 + (\psi_2 - \eta) \ggamma_2 +
  (\psi_3 - \xi) \ggamma_3 + (e^{(1)} + \tfrac{1}{2 \A_1}) C_1 +
  (e^{(2)} + \tfrac{1}{2 \A_2}) C_2 +
  e^{(3)} C_3 \\*
  {}+ \psi_0 \ggamma_0 - 2 \A_1 B_1 \ggamma_2 \ggamma_3 - 2 \A_2 B_2
  \ggamma_3 \ggamma_1 - 2 \A_3 B_3 \ggamma_1 \ggamma_2 - \bbeta_0
  \ggamma_1 C_1 - \bbeta_0 \ggamma_2 C_2 - \bbeta_0 \ggamma_3 C_3,
\end{multline}
where we have introduced the ghosts---bosonic and fermionic
first-order systems with the respective operator product expansions
\begin{equation}
  \bbeta_i(z)\,\ggamma_j(w)=\smash{\frac{-\delta_{ij}}{z-w}},
  \qquad
  B_i(z)\,C_j(w)=\smash[t]{\frac{\delta_{ij}}{z-w}}.
\end{equation}

Because the $\cA$ algebra involves the ``auxiliary'' $\eta,\xi$
system, the cohomology of the BRST operator must contain a Heisenberg
algebra.  The current generating this algebra is readily found as
\begin{equation}
  \widetilde{H} = 2 h^{(3)} + 2B_3 C_3
  + \bbeta_0 \ggamma_0 + \bbeta_1 \ggamma_1
  + \bbeta_2 \ggamma_2 - \bbeta_3 \ggamma_3 + \eta \xi.
\end{equation}
Our aim is to find a $W$ algebra in the ghost-number zero cohomology
of $\cQ$ commuting with this Heisenberg algebra~$\cH_0$.

To this end, we use the standard procedure of combining the Sugawara
energy-momentum tensor~$\cT$ (see Eq.~\eqref{D-sug}) ``improved'' by
 derivatives of the Cartan currents with the ghost energy-momentum
tensors.  This gives the family of energy-momentum tensors in the
cohomology of~$\cQ$ in the ghost number zero,
\begin{multline}
  \widetilde{T}_{(j)} = \cT + \d B_1 C_1 + \d B_2 C_2 + (2j-3) B_3 \d
  C_3 + 2(j-1) \d B_3 C_3+{}\\{} + (j-1) \bbeta_0 \d \ggamma_0 + j\d
  \bbeta_0 \ggamma_0 + (j-2) \bbeta_1 \d \ggamma_1 +
  (j-1) \d \bbeta_1 \ggamma_1
  + (j-2) \bbeta_2 \d \ggamma_2 + (j-1) \d \bbeta_2 \ggamma_2+{}\\
  {} + (1 - j)\bbeta_3 \d \ggamma_3 + (2 - j)\d \bbeta_3 \ggamma_3
  + (j-2) \eta \d \xi + (j-1) \d \eta \xi\\*
  {} + \d h^{(1)} + \d h^{(2)} + 2(j-1) \d h^{(3)},
\end{multline}
where $j$ is arbitrary.  We next construct a unique combination that
commutes with $\widetilde{H}$ (and is independent of $j$),
\begin{equation}
  \widehat{T} = \widetilde{T}_{(j)}
  - \frac{\A_3}{2(\A_3 + 2)} \widetilde{H}\widetilde{H} -
  \frac{4(j-1) + (2j - 3)\A_3}{2(\A_3 + 2)}\d\widetilde{H}.
\end{equation}
\begin{Lemma}
  The central charge of $\;\widehat{T}$ is given by
  \begin{equation}
    \widehat{c}  = 
    \frac{3 (\A_1-2) (\A_2-2) \A_3}{\A_1 \A_2 (\A_3 + 2)}.
  \end{equation}
  With the identifications
  \begin{equation}
    \A_1 = k_1 + 2,\qquad
    \A_2 = k_2 + 2, \qquad
    \A_3 = -k_1 - k_2 - 4,
  \end{equation}
  this becomes the central charge of the coset theory,
  Eq.~\eqref{the-c}.
\end{Lemma}

\noindent
Obviously, $\A_1$ and $\A_2$ can be transposed in the above formulas.
We also recall (see Appendix~\ref{app:D-all}) that the $\alpha$
parameter of $\hD$ is determined as $\alpha=-1-\frac{\A_i}{\A_j}$; we
can assume
\begin{equation}
  \alpha=-1-\frac{k_1+2}{k_2+2},
\end{equation}
which can also be replaced by any expression obtained by transposing
any two among the three levels $k_1$, $k_2$, and~$k_3=-k_1 - k_2 -4$.
We note that relation~\eqref{alpha-semi-classical} is a
``semiclassical'' (i.e., $k_1, k_2 \to\infty$) limit of one of these
formulae.

Thus, the cohomology of $\cQ$ contains, in addition to $\cH_0$, an
algebra that contains the Virasoro algebra with the central charge of
the coset theory $\hSL2_{k_1}\oplus\hSL2_{k_2}/\hSL2_{k_1 + k_2}$.  To
show that this extends to the entire $\WW\!D_{2|1}(k_1,k_2)$ algebra
commuting with $\cH_0$, we introduce a filtration on the BRST complex
such that
\begin{equation}\label{Q-decompose}
  \cQ=\cQ^{(0)} +  \cQ^{(1)} + \cQ^{(2)},
\end{equation}
with $\cQ^{(i)}$ decreasing the filtration index by~$i$ and
$(\cQ^{(0)})^2=0$, $\cQ^{(0)}\cQ^{(1)} + \cQ^{(1)}\cQ^{(0)}=0$, and
$(\cQ^{(1)})^2 + \cQ^{(0)}\cQ^{(2)} + \cQ^{(2)}\cQ^{(0)} = 0$.  The
respective currents are given by
\begin{multline}
  \cJ^{(0)} = \psi_1 \ggamma_1 + \psi_2 \ggamma_2 + \psi_3 \ggamma_3 +
  e^{(1)} C_1 + e^{(2)} C_2 + e^{(3)} C_3
  + \psi_0 \ggamma_0 \\*
  {} - 2 \A_1 B_1 \ggamma_2 \ggamma_3 - 2 \A_2 B_2 \ggamma_3 \ggamma_1
  - 2 \A_3 B_3 \ggamma_1 \ggamma_2 - \bbeta_0 \ggamma_1 C_1 - \bbeta_0
  \ggamma_2 C_2 - \bbeta_0 \ggamma_3 C_3,
\end{multline}
\begin{gather}
  \cJ^{(1)} =
  -\ggamma_1\eta - \ggamma_2\eta - \ggamma_3\xi,\\
  \cJ^{(2)} = \tfrac{1}{2 \A_1}C_1 + \tfrac{1}{2 \A_2}C_2.
\end{gather}
The second and the third terms in~\eqref{Q-decompose} can be viewed as
``perturbations'' of~$\cQ^{(0)}$, and $\cQ$ as a deformation
of~$\cQ^{(0)}$.  We now evaluate the spectral sequence associated with
this decomposition of the BRST differential.

The cohomology of $\cQ^{(0)}$ in the ghost number zero is generated by
$\eta\xi$ and three more currents
\begin{align}
  \widehat h^{(1)}&= h^{(1)} + B_1C_1 + \thalf \bbeta_0\ggamma_0 -
  \thalf \bbeta_1 \ggamma_1 + \thalf \bbeta_2 \ggamma_2 +
  \thalf \bbeta_3 \ggamma_3,\\
  \widehat h^{(2)}&= h^{(2)} + B_2C_2 + \thalf \bbeta_0\ggamma_0 +
  \thalf \bbeta_1 \ggamma_1 - \thalf \bbeta_2 \ggamma_2 +
  \thalf \bbeta_3 \ggamma_3,\\
  \widehat h^{(3)}&= h^{(3)} + B_3C_3 + \thalf \bbeta_0\ggamma_0 +
  \thalf \bbeta_1 \ggamma_1 + \thalf \bbeta_2 \ggamma_2 - \thalf
  \bbeta_3 \ggamma_3.
\end{align}
The Heisenberg algebra generated by the four currents gives the
ghost-number-zero part of the zeroth term in the spectral sequence.
To find the first term, we note the relation
\begin{multline}
  \cQ^{(0)}(\thalf\bbeta_2 C_3 \eta + \thalf\bbeta_3 C_2 \eta - \A_1
  \ggamma_0 B_1 \eta - \thalf\psi(\p, \m, \m)\eta)\\*
  -(\A_1 \widehat h^{(1)} - \A_2 \widehat h^{(2)} - (\A_3 + 2)
  \widehat h^{(3)} + \widetilde H)\ggamma_1\eta = -\d(\ggamma_1\eta).
\end{multline}
On the cohomology of $\cQ^{(0)}$, this becomes a homogeneous
differential equation, and therefore, the first term in $\cQ^{(1)}$
acts on the cohomology of $\cQ^{(0)}$ as a vertex operator,
\begin{align}
  \ggamma_1\eta&=e^{\int(\A_1 \widehat h^{(1)} - \A_2 \widehat h^{(2)}
    - (\A_3 + 2) \widehat h^{(3)}
    + \widetilde H)}.\\
  \intertext{For the other terms in $\cQ^{(1)}$, we similarly find
    that on the cohomology of $\cQ^{(0)}$, they are given by}
  \ggamma_2\eta&=e^{\int(-\A_1\widehat h^{(1)} + \A_2\widehat h^{(2)}
    -(\A_3 + 2)\widehat h^{(3)}
    + \widetilde H)},\\
  \ggamma_3\xi&=e^{\int(-\A_1\widehat h^{(1)} -\A_2\widehat h^{(2)}
    +(\A_3+2)\widehat h^{(3)} -\widetilde H)}.
\end{align}

With the exponents in the last three formulae denoted as $X_a$,
$a=1,2,3$, we have
\begin{align}
  X_1(z)X_2(w)&=(\A_3 + 1)\log(z-w),\\
  X_1(z)X_3(w)&=(\A_2 - 1)\log(z-w),\\
  X_2(z)X_3(w)&=(\A_1 - 1)\log(z-w).
\end{align}
With the scalar product determined by the operator products, we then
have $\langle X_1, X_2\rangle + \langle X_1, X_3\rangle + \langle X_2,
X_3\rangle=-1$.  In addition, $\langle X_a, X_a\rangle=1$ because
\begin{equation}
  X_a(z)X_a(w)=\log(z-w),
\end{equation}
and therefore, \textit{the three operators $\ggamma_1\eta$,
  $\ggamma_2\eta$, and $\ggamma_3\xi$ acting on the ghost-number-zero
  part of the zeroth term of the spectral sequence are represented by
  fermionic screenings that are equivalent to those in
  Eqs.~\eqref{sigma-1-final}--\eqref{sigma-3-final}}.  The precise
identification involves splitting the space of the four currents
$(\eta\xi, \d X_1, \d X_2, \d X_3)$ into the one-dimensional space
spanned by $\widetilde H$ and the orthogonal complement; acting on the
latter, the operators $\ggamma_1\eta$, $\ggamma_2\eta$, and
$\ggamma_3\xi$ single out the $W$ algebra $\WW\!\D$.

Thus, the first term of the spectral sequence contains the $W$ algebra
$\WW\!\D$ (which was defined as the vertex operator algebra in the
commutant of the fermionic screenings in
Eqs.~\eqref{sigma-1-final}--\eqref{sigma-3-final}).
Following~\cite{[FFr]}, one shows that for generic values of the
parameters, the cohomology of~$\cQ$ is precisely this $W$ algebra
times $\cH_0$ (in particular, $\cQ^{(2)}$ is trivial on the first term
of the spectral sequence).  This completes the demonstration of the
Hamiltonian reduction $\hD\to\WW\!\D$.

\section{Discussion and conclusions} \label{sec:discussion}
We now briefly discuss interesting alternative constructions of the
$W$ algebra $\WW\!\D$ corresponding to the coset
$\hSL2_{k_1}\oplus\hSL2_{k_2}/\hSL2_{k_1 + k_2}$ and note several
points for the future development.

\subsection{$\WW D(2|1;\alpha)$ from {\sc bfb} screenings}
\label{BFB-scr} The algebra $\D$ 
has another maximal nilpotent subalgebra; it corresponds to the simple
root system consisting of one fermionic and two bosonic roots.  This
gives another possibility for the Hamiltonian reduction of $\hD$.  The
result is a $W$ algebra belonging to the same family
$\WW\!D_{2|1}(k_1,k_2)$, although the values of $k_1$ and $k_2$ are
generically different from those resulting from the reduction in
Sec.~\ref{sec:Ham-red}.

The fermionic root system corresponds, in a certain sense, to viewing
$\D$ as three $\SSL21$ algebras (with the fermionic simple roots) that
pairwise intersect over $\SSL11$.  At the same time, recalling that
$\SSL21$ admits the simple root system consisting of one bosonic and
one fermionic root, we can view $\D$ as \textit{two} $\SSL21$ algebras
(intersecting again over~$\SSL11$).  The corresponding Chevalley
generators of $\D$ are $\psi$ (a fermion) and $e_1$ and $e_2$ (bosons)
that satisfy, in addition to $[\psi,\psi]=0$, the Serre relations
\begin{equation*}
  [e_1, e_2]=0,\quad
  [e_1, [e_1, \psi]]=0,\quad
  [e_2, [e_2, \psi]]=0
\end{equation*}
(the pairs $(e_1,\psi)$ and $(e_2,\psi)$ are the simple root systems
of the two $\SSL21$ algebras).

The description of $\WW\!\D$ as the commutant of one fermionic and two
bosonic screenings can be deduced from the ``deformation'' picture in
Sec.~\ref{sec:constr-coset-deform}, where we now take the $\hSL2_k$
algebra \textit{in the standard (asymmetric) bosonization} described
in Sec.~\ref{subsec:standard-realisation}.  There then exist the
fermionic screening $\sigma_{\mathrm{F}}=\oint e^{f}$ and the Dotsenko
screening
\begin{equation}
  \sigma_{\mathrm{D}} = \oint\betaW^{-k-2}\,
  e^{\sqrt{2(k+2)}\varphi},
\end{equation}
where \textit{the integrand is also an exponential}, with non-integral
powers of $\betaW$ well-defined in terms of the bosonization in
Eq.~\eqref{beta-gamma-bos} (this screening does not exist in the
bosonization used in Sec~\ref{sec:sym-bos}).  The
$\sigma_{\mathrm{D}}$ screening and the standard Wakimoto bosonization
screening $\sigma_{\mathrm{W}}$ make up a \textit{Virasoro pair},
i.e., the respective integrands are given by
$e^{\Vec\mu\cdot\Vec\Phi}$ and $e^{\Vec\mu'\cdot\Vec\Phi}$ (with
$\Vec\Phi=(\phi, \varphi,f)$), where $\Vec\mu' =
\frac{-2}{\Vec\mu^2}\Vec\mu$.  The commutant of $(\sigma_{\mathrm{W}},
\sigma_F, \sigma_{\mathrm{D}})$ is the $\hSL2$ algebra.  The pair
$(\sigma_F, \sigma_{\mathrm{D}})$ corresponds to the $\SSL21$ simple
root system and, thus, generates the nilpotent subalgebra
of~$\uU_q\SSL21$.

Next, the standard bosonization allows constructing the
$\widetilde{\sigma}^+$ screening that makes a Virasoro pair
with~$\sigma^+$.  It also \textit{commutes with the commutant of the
  other screenings}.  The three screenings
\begin{equation}\label{bfb-screening}
  \sigma_{\mathrm{D}},\qquad \sigma_{\mathrm{F}},\qquad
  \widetilde{\sigma}^+
\end{equation}
(all of which have purely exponential integrands) generate the
nilpotent subalgebra of $\uU_q\D$ corresponding to the simple root
system consisting of one fermionic and two bosonic roots and can
therefore be used to single out the $\WW\!\D$ algebra.  We note that
$(\sigma_F, \widetilde{\sigma}^+)$ corresponds to the simple root
system of the second $\uU_q\SSL21$ quantum group.

It is instructive to identify the bosonic screenings corresponding to
the three $\uU_q\SL2$ quantum groups in this approach.  These are
$\sigma^-$ (which is expressed as the integral
of~\eqref{minus-integrand}, a somewhat unconventional representation
for a screening), $\sigma_{\mathrm{W}}$ (the standard Wakimoto
screening), and~$\sigma^+$.

Thus, with the ``\textsc{bfb}'' simple root system, we have used the
symmetric bosonization to reproduce the quantum group content of the
coset theory, the mutually commuting quantum groups $\uU_q\D$ and
$\uU\SL2_{q'}\tensor\uU\SL2_{q''}\tensor\uU\SL2_{q'''}$.

\subsection{Matter realization of $\WW D(2|1;\alpha)$}
\label{subsec:matter}  $W$ algebras are 
vertex operator algebras selected by a set of screenings in a
free-field theory.  In this sense, $W$ algebras are deformations of
free-field theories with the help of operators that are
\textit{primary fields} in these theories.  A natural generalization
is to take a more general conformal field theory and deform it using
some of its primary fields as screenings.

The $\WW\!\D$ algebra can be realized in this spirit as a subalgebra
in the product of two (universal enveloping of) Virasoro algebras and
a Heisenberg algebra.  We somewhat loosely refer to this system as
``matter'' (consisting of the Virasoro algebras with the respective
central charges $d_1$ and $d_2$) dressed with a free field $(\phi)$.
It can be arrived at by describing $\WW\!\D$ as the commutant of the
screenings in~\eqref{bfb-screening} and first evaluating the kernels
of the two bosonic screenings, which gives two Virasoro algebras and a
free scalar field.  To construct $\WW\!\D$ from
$\Vir(d_1)\oplus\Vir(d_2)\oplus(\phi)$, we must then use the analogue
of the fermionic screening in this theory.  This requires generalizing
the concept of the fermionic screenings (which so far have only been
defined in free-field representations).

We first recall a similar construction of the $W$ algebra $\WW\SSL21$
as a deformation of the product of a single Virasoro algebra and a
free field $\phi$ with the help of the $\Phi_{12}$ operator.  This
vertex operator furnishes a two-dimensional representation of the
quantum group $\uU_q\SL2$; we let $\Phi_{12}^{\pm}$ denote the two
components.  The operator can be dressed with a scalar such that
$\Phi_{12}^{+}\,e^{\zeta\phi}$ becomes a fermionic operator in the
sense of the operator product $\Phi_{12}^{+}\,e^{\zeta\phi}(z)\cdot
\Phi_{12}^{+}\,e^{\zeta\phi}(w)\sim(z-w)$; it then follows that
$\Phi_{12}^{-}$ also is a fermionic operator.  It follows that the
commutant of $\oint\Phi_{12}^+ e^{\zeta\phi}$ and $\oint\Phi_{12}^-
e^{\zeta\phi}$ is the $\WW\SSL21$ algebra.

In this construction of $\WW\SSL21$, the quantum group $\uU_q\SL2$
associated with the Virasoro representation theory is extended by its
two-dimensional representation $\oC_q^2$ (as before, the subscript
indicates that this is a quantum group representation---the one on
$\Phi_{12}^{+}$ and $\Phi_{12}^{-}$---but we omit it for brevity in
what follows).  We observe that
\begin{equation}
  \uU_q\SL2 \oplus \oC^2
\end{equation}
is a part of the $\uU_q\SSL21$ quantum group, whose generators can be
arranged as
\begin{equation}
  \oC^2 \oplus
  \!\begin{array}{c}
    \uU_q\SL2\\
    \oplus\\
    \oC
  \end{array}\!
  \oplus\oC^2
\end{equation}
This extension of $\uU_q\SL2$ to $\uU_q\SSL21$ is indeed known in the
$\hSL2$ representation theory~\cite{[FM]}.

Similarly to this construction of $\WW\SSL21$, the $\WW\!\D$ algebra
is the subalgebra in the tensor product of (the universal enveloping
algebras of) two Virasoro algebras and a Heisenberg algebra that
commutes with the integral of the product of the respective
$\Phi_{12}$ fields dressed with a scalar,
\begin{equation}\label{sS}
  \sS = e^{\zeta\phi}\,\Phi^{(1)}_{12}\,\Phi^{(2)}_{12}.
\end{equation}
This operator is represented by $(\oC^2\tensor\oC^2)(z)$.  The $\zeta$
parameter is chosen such that the
$e^{\zeta\phi}{\Phi^{(1)}_{12}}^+\!{\Phi^{(2)}_{12}}^+$ component be a
fermion (with the central charges expressed as $d_i=13-6(k_i+2) -
\tfrac{6}{k_i+2}$, $i=1,2$, and with $\phi(z)\phi(w)=-\log(z-w)$, this
fixes $\zeta^2 = \half(k_1 + k_2 + 2)$).  The commutant of $\oint
dz\sS(z)$ on the product of the two Virasoro algebras and the
free-field theory is the $W$ algebra $\WW\!\D$.  This gives the
``matter'' construction of the $\hSL2\oplus\hSL2/\hSL2$ coset.

At the quantum group level, the appearance of $\uU_q\D$ in the above
construction can be seen if its generators are arranged as
\begin{equation}\label{Dq-arrange}
  \uU_q\D\supset
  \oC\oplus\oC^2\tensor\oC^2 \oplus
  \!\begin{array}{c}
    \uU_{q'}\SL2\oplus\uU_{q''}\SL2\\
    \oplus\\
    \oC
  \end{array}\!
  \oplus\oC^2\tensor\oC^2\oplus\oC
\end{equation}
(where $q'$ and $q''$ are combined into the $q$ and $\alpha$
parameters of~$\uU_q\D$, which therefore depends on two parameters;
classically, the $\D$ fermions $\oC^2\tensor\oC^2\tensor\oC^2$ are
broken to $\oC^2\tensor\oC^2\oplus\oC^2\tensor\oC^2$ and one of the
$\SL2$ subalgebras to $\oC\oplus\oC\oplus\oC$).  In the
$\uU_{q'}\SL2\oplus\uU_{q''}\SL2\oplus\oC^2\tensor\oC^2$ part
in~\eqref{Dq-arrange}, we now interpret
$\uU_{q'}\SL2\oplus\uU_{q''}\SL2$ as the quantum group symmetry of
$\Vir(d_1)\oplus\Vir(d_2)$, and $\oC^2\tensor\oC^2$ as the $\oint\sS$
operator.

\subsection{Quantum group representations} A manifestation
of the quantum-group symmetry of conformal field theories is the
correspondence between primary fields of vertex operator algebras and
quantum group representations.  The fusion of primary fields then
corresponds to tensor products of quantum group representations.  The
existence of a class of quantum group representations (for example,
infinite-dimensional ones) that are closed under direct sums and
tensor products allows one to select a class of vertex operator
algebra representations that are closed under operator products.  For
example, the Virasoro primary fields $\Phi_{mn}$ are thus determined
by the $m$- and $n$-dimensional representations of two $\uU_q\SL2$
quantum groups with the respective $q$ parameters determined by
$\alpha_+$ and $\alpha_-$; these two quantum $\SL2$ groups commute
with each other.

Primary fields of $\WW\!D_{2|1}(k_1,k_2)$ can be labeled as
\begin{equation*}
  \Psi_{\nu;n_1,n_2,n_3}
\end{equation*}
where $\nu$ is a $\uU_q\D$ (multi)index and $n_i$ are labels of
$\uU_{q_1}\SL2\tensor\uU_{q_2}\SL2\tensor\uU_{q_3}\SL2$ with the
respective quantum group parameters $q_j=e^{\pi i\kappa_j}$,
see~\eqref{the-kappa}.  For example, the vertex operators used in
reconstructing $\hSL2_{k_1}$ and $\hSL2_{k_2}$ are
$\Psi_{212}=\Psi_{1;2,1,2}$ and $\Psi_{122}=\Psi_{1;1,2,2}$, as we saw
in Sec.~\ref{sec:qg-doublets}.  These ``lowest'' operators are similar
to $\Phi_{21}$ for the Virasoro algebra, which are singlets with
respect to one of the $\uU_q\SL2$ quantum groups and doublets with
respect to the other.

The two quantum groups
$\uU_{q_1}\SL2\tensor\uU_{q_2}\SL2\tensor\uU_{q_3}\SL2$ and $\uU_q\D$
commute with each other.  We note that while the presence of three
$\uU_q\SL2$ quantum groups is rather natural once the $\WW\!\D$
algebra is interpreted as the $\hSL2_{k_1}\oplus\hSL2_{k_2}/\hSL2_{k_1
  + k_2}$ coset, the emergence of $\uU_q\D$ is unexpected (and
somewhat mysterious) from the \textit{coset} point of view.

\subsection{Resolutions and characters} A further study of
representations of the relevant quantum groups can lead to the
construction of \textit{resolutions} (and hence, character formulae)
for $\WW\!D_{2|1}(k_1,k_2)$ representations.  We concentrate on the
vacuum representations, which are interesting because they describe
the field content of the theory.  For generic $k_1$ and $k_2$, the
character formula can be found by analyzing kernels of the screenings.
Using the ``deformation'' picture in
Sec.~\ref{sec:constr-coset-deform}, one first establishes that the
same $W$ algebra $\WW\!D_{2|1}(k_1,k_2)$ is the commutant of one
bosonic and two fermionic screenings provided these are chosen such
that the momentum of the bosonic screening is not proportional to the
sum of the two fermionic screening momenta.  The commutant of the
fermionic screenings is the $\hSL2$ algebra, and hence,
$\WW\!D_{2|1}(k_1,k_2)$ is the kernel of $\sigma^+=\oint
J^+\,e^{\beta\int J^0}$ 
on the weight-zero subalgebra of $\uU\hSL2$ (where the weight is the
eigenvalue $j$ of $J^0_0$, the zero mode of the Cartan current).  The
vacuum representation of~$\WW\!D_{2|1}(k_1,k_2)$ can therefore be
found as the kernel of $\sigma^+$ in the weight-zero subspace of the
vacuum $\hSL2$ representation.  By the deformation argument (taking
$\beta\to0$) for generic~$k_1$ and~$k_2$, the character of this space
is the same as the character of the kernel of $J^+_0=\oint J^+(z)$ in
the weight-zero subspace of the vacuum $\hSL2$ representation (in
other words, the space of \hbox{invariants of the $\SL2$ action in the
  vacuum $\hSL2$ representation}).

Because the mapping via $J^+_0$ is an epimorphism of the weight-$0$
subspace on the weight-$1$ subspace in the vacuum representation, the
character of the kernel is given by the difference of the characters
of the weight-$0$ and weight-$1$ subspaces,
\begin{equation}
  \chi^{\phantom{y}}_{\Vac_{\WW\!\D}}(q) = \chi_{j=0}(q) -
  \chi_{j=1}(q).
\end{equation}
Next, $\chi_{j=0}(q)$ can be found by either studying the resolution
associated with the fermionic screening, with the result
\begin{equation}\label{char-0}
  \chi_{j=0}(q) =
  \smash[t]{\frac{1 + 2 \sum_{n=1}^{\infty}(-1)^n q^{\half n(n+1)}}{
      \prod_{i=1}^\infty(1 - q^i)^{3}}},
\end{equation}
or directly taking the $z^0$-component of the vacuum $\hSL2$ character
\begin{equation}\label{vac-sl2-character}
  \frac{1}{\prod_{i=1}^\infty(1 - z q^i)\,
    \prod_{i=1}^\infty(1 - z^{-1}q^i)\,
    \prod_{i=1}^\infty(1 - q^i)}
\end{equation}
(which reproduces~\eqref{char-0}).  Similarly, the resolution gives
\begin{equation}
  \chi_{j=1}(q) =
  \frac{\sum_{n=1}^{\infty}(-1)^{n+1} q^{\half n(n+1)}
    + \sum_{n=2}^{\infty}(-1)^{n+1} q^{\half n(n+1) - 1}}{
    \prod_{i=1}^\infty(1 - q^i)^{3}},
\end{equation}
which also is the $z^1$-component of~\eqref{vac-sl2-character}.

The character of the vacuum representation of the $W$ algebra is
therefore given by
\begin{multline}
  \chi^{\phantom{y}}_{\Vac_{\WW\!\D}}(q)= \frac{1 + 3
    \sum_{n=1}^{\infty}(-1)^n q^{\half n(n+1)} +
    \sum_{n=2}^{\infty}(-1)^{n} q^{\half n(n+1) - 1}}{
    \prod_{i=1}^\infty(1 - q^i)^3}=\\
  =1 + q^2 + q^3 + 3 q^4 + 3 q^5 + 8 q^6 + 9 q^7 + 19 q^8 + 25 q^9 +
  45 q^{10} + 61 q^{11} + 105 q^{12} + 144 q^{13}
  + \dots
\end{multline}
The coefficient $3$ at $q^4$ corresponds to the two dimension-$4$
descendants of the energy-momentum tensor and the~$F$ field found in
Appendix~\ref{app:formula}.

This character formula must also follow from a resolution constructed
using three fermionic screenings.  This situation is standard in that
this resolution of $\WW\D$-modules has the same structure as the
resolution of the trivial $\uU_q\D$ representation via $\uU_q\D$ Verma
modules, which is given by
\begin{equation}\dgARROWLENGTH=1.5em
  \dots
  \begin{diagram}
    \node{}\arrow{e,t}{} \node{
      \begin{matrix}
        \bullet\\ \bullet\\ \bullet\\ \bullet
      \end{matrix}
      }\arrow{e,t}{} \node{
      \begin{matrix}
        \bullet\\ \bullet\\ \bullet\\ \bullet
      \end{matrix}
      }\arrow{e,t}{} \node{
      \begin{matrix}
        \bullet\\ \bullet\\ \bullet\\ \bullet
      \end{matrix}
      }\arrow{e,t}{} \node{
      \begin{matrix}
        \bullet\\ \bullet\\ \bullet
      \end{matrix}
      }\arrow{e,t}{} \node{\bullet} \arrow{e,t}{} \node{\bullet}
  \end{diagram}
\end{equation}
This shows that the number of modules at each term stabilizes at~$4$
(cf.\ the resolutions with a growing number of modules in each
term~\cite{[FSST]} and resolutions of a more exotic
shape~\cite{[b-fly]}), but the details of the mappings realized by the
fermionic screenings are left for a future work.

We note that there are a number of interesting ``degenerate'' cases
corresponding to special values of $k_1$ and $k_2$ (see
also~\cite{[Sch]} and references therein.).  For example, a
Felder-like complex arises if the $\sigma^+$ screening can be applied
repeatedly, e.g.,
\begin{equation}\dgARROWLENGTH=1.6em
  \dots\kern-8pt
  \begin{diagram}
    \node{}\arrow{e,t}{\sigma^+}
    \node{\bullet}\arrow{e,t}{(\sigma^+)^N}
    \node{\bullet}\arrow{e,t}{\sigma^+}
    \node{\bullet}\arrow{e,t}{(\sigma^+)^N} \node{}
  \end{diagram}
  \kern-8pt\dots
\end{equation}
with $(\sigma^+)^{N+1}=0$.  The arrows, obviously, map between
subspaces with different $J^0_0$ eigenvalues.  The cohomology of this
complex gives the vacuum representation of the coset for the
corresponding special values of $k_1$ and $k_2$.  These parameter
values can then allow taking $\hSL2$ representations other than the
vacuum one and constructing resolutions using the corresponding weight
spaces (e.g., for positive integer $k_1$ and rational $k_2$);
comparing these would result in nontrivial character identities.
There also exist numerous possibilities for arranging a Felder-like
complex along the directions corresponding to other screenings.

\bigskip

\noindent
\textbf{6.5.  Other remarks.}  The same $W$ algebra can be obtained by
reducing $\hD$ with different parameters; represen\-tation-theory
implications of this fact may be worth being investigated.  For a
fixed $\hD$ algebra, on the other hand, the $\xi$ and $\eta$ fields
can be used in the reduction constraints (cf.
Eqs.~\eqref{the-constraints}) in three different ways, which gives
\hbox{different $W$ algebras from the same family},
\begin{equation}\dgARROWLENGTH=.5em
  \begin{diagram}
    \node[2]{\hD}\arrow{sse, t}{}
    \arrow{ssw, t}{}\\[2]
    \node{\WW\!D_{2|1}(k_1,k_2)} \arrow[2]{e,t,..}{}
    \node[2]{\WW\!D_{2|1}(k'_1,k'_2)}\arrow[2]{w,b,..}{}
  \end{diagram}
\end{equation}
where the dotted line denotes a \textit{correspondence} between
(representation theories of) the two algebras, which can be interesting
to explore.

The construction of $\WW\!\D$ in terms of two Virasoro algebras
dressed with a scalar suggests that the $\hD$ algebra can be realized
in terms of two Virasoro algebras or their supersymmetric extensions
and several more free fields, similarly to the realizations of
$\hSL2$~\cite{[S-sing]} and $\widehat{osp}(1|2)$ and
$\hSSL21$~\cite{[S-sl21]}.

Because a partial Hamiltonian reduction of $\hD$ is related to the
$\N4$ superconformal algebra~\cite{[IM],[IMP]}, this algebra must be
related to the $\hSL2\oplus\hSL2/\hSL2$ coset via a ``secondary''
Hamiltonian reduction~\cite{[MR]}.

Finally, we note again that the emergence of $\uU_q\D$ as a part of
``quantum symmetries'' of $\WW\!D_{2|1}(k_1,k_2)$ is rather mysterious
when this $W$ algebra is viewed as the coset conformal field theory.
An interesting problem is to construct the
$\hSL2_{k_1}\oplus\hSL2_{k_2}/\hSL2_{k_1 + k_2}$ vertex operators that
carry nontrivial $\uU_q\D$ representations.

\bigskip

\noindent
\textbf{Acknowledgments}.  We are grateful to A.~Taormina for helpful
remarks on $\N4$ algebras and for drawing our attention to works on
the $\hD\to(\N4)$ reduction.  A part of the paper was written when the
authors were visiting the Fields Institute, and the kind hospitality
extended to us there is gratefully acknowledged.  This paper was
supported in part by the RFBR Grant~01-01-00906 and the Russian
Federation President Grant~99-15-96037, and also by INTAS-OPEN-97-1312
and RFBR Grant~99-01-01169.

\appendix
\section{The dimension-four field} \label{app:formula} The commutant
of the screenings $\sigma_i$ constructed in Eqs.~\eqref{sigma-i},
\eqref{S-i}, and~\eqref{a-found1}--\eqref{a-found3} contains
energy-momentum tensor~\eqref{T-become} and possibly, higher-dimension
fields generating a $W$ algebra.  We now assume $k_1$ and $k_2$ to be
generic.  For $n=-1$ and only for $n=-1$, the commutant contains a
primary dimension-4 field $F$.  Using~\cite{[Th]}, this can be found
as
\begin{multline}\label{eq:F}
  \xi^{-1}\,F= \tfrac{K^2}{\sqrt{2 k_2} (k_2 + 2)} 
  \sqrt{k_1} (k_1 + 2)^2 (k_2 - 1)
  (3 k_2 + 4) (3 k_2 + 11) 
  (\d\varphi_3)^4
  \\{}- 
  \tfrac{2 K^{3/2}}{(k_2 + 2)}
  \sqrt{k_1}  (k_1 + 2)^{3/2} (k_2 - 1)
  (3 k_2 + 4)  (3 k_2 + 11)
  \d^2\varphi_3(\d\varphi_3)^2  
  \\{}+ 
  \tfrac{3\sqrt{k_1}}{2\sqrt{2 k_2}} 
  (k_1 + 2) (3 k_2 + 4) 
  (41 k_2 k_1^2 + 22 k_1^2 + 41 k_2^2 k_1 + 186 k_2 k_1 + 
  88 k_1 + 52 k_2^2 + 168 k_2 + 88)\times{}\\*
  \mbox{} \hfill{}\times K \bigl[
  (\d\varphi_1)^2(\d\varphi_2)^2
  -
  \d^2\varphi_1(\d\varphi_2)^2
  \bigr]
  \\{}
  + \tfrac{3\sqrt{k_1}}{\sqrt{2k_2}} 
  (k_1 + 2) (3 k_2 + 4) 
  (2 k_2 k_1^2 - 11 k_1^2 + 2 k_2^2 k_1 - 3 k_2 k_1 - 44 k_1
  + 4 k_2^2 - 4 k_2 - 44)\times{}\\
  \mbox{}\hfill{}\times K
  \bigl[(\d\varphi_1)^2(\d\varphi_3)^2
  + 
  (\d\varphi_2)^2(\d\varphi_3)^2
  -
  \d^2\varphi_1(\d\varphi_3)^2
  \bigr]
  \\{}
  + \tfrac{3  i\sqrt{k_1}}{\sqrt{2(k_2 + 2)}}
    (k_1 + 2)
  (3 k_2 + 4)
  (2 k_2 k_1^2 - 11 k_1^2 + 2 k_2^2 k_1 - 3 k_2 k_1 - 44 k_1
  + 4 k_2^2 - 4 k_2 - 44)K    
  \d^2\varphi_2(\d\varphi_3)^2  
  \\{}+
  \tfrac{\sqrt{k_1}}{2\sqrt{2}(k_2 + 2)} 
  (k_1 + 2)  (k_2 - 1)
  \sqrt{k_2}  (3 k_2 + 4)  
  (15 k_1^2 + 15 k_2 k_1 + 60 k_1 + 24 k_2 + 68) K
  \d^2\varphi_3\d^2\varphi_3  
  \\{}-
  \tfrac{\sqrt{k_1}}{\sqrt{2}} (k_1 - 1)  
  (3 k_1 + 4)\sqrt{k_2}  
  (5 k_2^2 + 5 k_1 k_2 + 20 k_2 + 4 k_1 + 8) K
  \d^3\varphi_2\d\varphi_2  
  \\{}-
  \tfrac{\sqrt{k_1}}{\sqrt{2}(k_2 + 2)} 
  (k_1 + 2)  (k_2 - 1)
  \sqrt{k_2}  (3 k_2 + 4)  
  (5 k_1^2 + 5 k_2 k_1 + 20 k_1 + 4 k_2 + 8) K
  \d^3\varphi_3\d\varphi_3  
  \\{}
  - \tfrac{\sqrt{k_1}}{k_2 + 2} 
  \sqrt{k_1 + 2} (2 k_1 + k_2 + 4)(2 k_2 + 3)
  P
  \sqrt{K}
  (\d\varphi_1)^3\d\varphi_3
  -
  \sqrt{k_1 (k_1 + 2)} (3 k_1 + 4)
  P
  \sqrt{K}
  \d^2\varphi_2\d\varphi_2\d\varphi_3  
  \\{}
  + \thalf\sqrt{k_1 (k_1 + 2)}  (3 k_2 + 4) 
  (49 k_2 k_1^2 - 22 k_1^2 + 49 k_2^2 k_1 + 174 k_2 k_1 - 88 k_1 + 
  68 k_2^2 + 152 k_2 - 88)\times{}\\
  \mbox{}\hfill{}\times\sqrt{K}
  \bigl[\d^2\varphi_1\d^2\varphi_3
  -
  (\d\varphi_1)^2\d^2\varphi_3
  -
  (\d\varphi_2)^2\d^2\varphi_3
  \bigr]
  \\{}  
  + \Bigl(\tfrac{i\sqrt{k_1}}{\sqrt{k_2}}
  \sqrt{k_1 + 2}\sqrt{k_2 + 2}  
  (\d\varphi_2)^3\d\varphi_3
  + \tfrac{3 i\sqrt{k_1(k_1 + 2)}}{\sqrt{k_2(k_2 + 2)}}
  (k_2 + 1)  
  (\d\varphi_1)^2\d\varphi_2\d\varphi_3\Bigr)
  (2 k_1 + k_2 + 4) P
  \sqrt{K}
  \\{}  
  + \sqrt{k_1 (k_1 \!+ 2)}  (3 k_1 \!+ 3 k_2 + 8)
  P
  \sqrt{K}
  \d^2\varphi_1\d\varphi_1\d\varphi_3
  - \tfrac{3 i\sqrt{k_1(k_1 + 2)}}{\sqrt{k_2(k_2 + 2)}}     
  (k_2 \!+ 1)  (2 k_1 \!+ k_2 + 4)
  P
  \sqrt{K}
  \d^2\varphi_1\d\varphi_2\d\varphi_3  
  \\{}  
  - \tfrac{i \sqrt{k_1 k_2}}{2\sqrt{k_2 + 2}} 
  \sqrt{k_1 \!+ 2} (3 k_2 \!+ 4)  
  (49 k_2 k_1^2 - 22 k_1^2 + 49 k_2^2 k_1 + 174 k_2 k_1 - 
  88 k_1 + 68 k_2^2 + 152 k_2 - 88)\sqrt{K}
  \d^2\varphi_2\d^2\varphi_3    
  \\{}
  - \tfrac{\sqrt{k_1 (k_1 + 2)}}{2 (k_2 + 2)}     
  (k_2^2 - k_1 k_2 + 3 k_2 + 4)
  P
  \sqrt{K}
  \d^3\varphi_1\d\varphi_3
  - \tfrac{i \sqrt{k_1 k_2}\sqrt{k_1 + 2}}{2\sqrt{k_2 + 2}}
  (k_1 - k_2)
  P
  \sqrt{K}
  \d^3\varphi_2\d\varphi_3  
  \\{}
  + \tfrac{\sqrt{k_1 (k_1 + 2)}}{12(k_2 + 2)}
  (k_2 - 1)  k_2  
  (3 k_2 + 4)  
  (5 k_1^2 + 5 k_2 k_1 + 20 k_1 + 4 k_2 + 8)\sqrt{K}
  \d^4\varphi_3 
  \\{}
  - \tfrac{\sqrt{k_1 k_2}}{4\sqrt{2}(k_2 + 2)}   
  (297 k_2^2 k_1^4 + 696 k_2 k_1^4 + 384 k_1^4 + 
  594 k_2^3 k_1^3 + 3768 k_2^2 k_1^3 + 6336 k_2 k_1^3 + 
  3072 k_1^3 + 297 k_2^4 k_1^2\\{} + 3768 k_2^3 k_1^2
  + 14256 k_2^2 k_1^2 + 20036 k_2 k_1^2
  + 9280 k_1^2 + 696 k_2^4 k_1 + 6336 k_2^3 k_1 + 20036 k_2^2 k_1\\{}
  + 26448 k_2 k_1 + 12544 k_1 + 384 k_2^4 + 3072 k_2^3 + 9280 k_2^2
  + 12544 k_2 + 6400) 
  (\d\varphi_1)^4
  \\{}+
  \tfrac{i\sqrt{2 k_1}}{\sqrt{k_2 + 2}}
  (k_2 + 1)
  (3 k_1^2 + 3 k_2 k_1 + 12 k_1 + 5 k_2 + 12)
  P
  (\d\varphi_1)^3\d\varphi_2
  \\{}+     
  \tfrac{i\sqrt{k_1} (k_1 + 2)}{2\sqrt{2(k_2 + 2)}}     
  (147 k_1 k_2^4 + 204 k_2^4 + 294 k_1^2 k_2^3 + 
  1142 k_1 k_2^3 + 1048 k_2^3 + 147 k_1^3 k_2^2 + 
  1142 k_1^2 k_2^2 + 2300 k_1 k_2^2 \\{}+ 1448 k_2^2 + 
  204 k_1^3 k_2 + 900 k_1^2 k_2 + 856 k_1 k_2 - 16 k_2 - 
  176 k_1^2 - 704 k_1 - 704)
  \bigl[(\d\varphi_1)^2\d^2\varphi_2
  -
  \d^2\varphi_1\d^2\varphi_2
  \bigr]
  \\{}
  + \tfrac{\sqrt{k_1}}{4\sqrt{2k_2}}
  (k_2 + 2) 
  (147 k_2 k_1^4 + 132 k_1^4 + 294 k_2^2 k_1^3 + 
  1440 k_2 k_1^3 + 1056 k_1^3 + 147 k_2^3 k_1^2 + 
  1880 k_2^2 k_1^2 + 4872 k_2 k_1^2 \\{}+ 3168 k_1^2 + 
  572 k_2^3 k_1 + 3752 k_2^2 k_1 + 6912 k_2 k_1 + 4224
  k_1 +  
  496 k_2^3 + 2336 k_2^2 + 3504 k_2 + 2112)  
  (\d\varphi_2)^4
  \\{}+     
  \tfrac{\sqrt{k_1 k_2}}{2\sqrt{2}(k_2 + 2)}     
  (75 k_2^2 k_1^4 + 210 k_2 k_1^4 + 120 k_1^4 + 
  150 k_2^3 k_1^3 + 1316 k_2^2 k_1^3 + 2716 k_2 k_1^3 + 
  1488 k_1^3 + 75 k_2^4 k_1^2\\{}
  + 1316 k_2^3 k_1^2 + 6504 k_2^2 k_1^2 + 11220 k_2 k_1^2 + 6112 k_1^2
  + 210 k_2^4 k_1 + 2716 k_2^3 k_1 + 11220 k_2^2 k_1\\{}
  + 18352 k_2 k_1
  + 10432 k_1 + 120 k_2^4 + 1488 k_2^3 + 6112 k_2^2 + 10432 k_2 + 6400)
  \d^2\varphi_1(\d\varphi_1)^2
  \\{}-
  \tfrac{i\sqrt{2k_1}}{\sqrt{k_2 + 2}} 
  (k_1 + 2)  
  (k_2 + 1)  (3 k_1 + 3 k_2 + 8)
  P
  \d^2\varphi_1\d\varphi_1\d\varphi_2
  \\{}+
  \tfrac{\sqrt{k_1 k_2}}{4\sqrt{2}(k_2 + 2)}    
  (237 k_2^2 k_1^4 + 528 k_2 k_1^4 + 288 k_1^4 + 
  474 k_2^3 k_1^3 + 2804 k_2^2 k_1^3 + 4328 k_2 k_1^3 + 
  1952 k_1^3 + 237 k_2^4 k_1^2\\{}
  + 2804 k_2^3 k_1^2 + 9632 k_2^2 k_1^2 + 11860 k_2 k_1^2
  + 4672 k_1^2 + 528 k_2^4 k_1 + 4328 k_2^3 k_1 + 11860 k_2^2 k_1\\{}
  + 12752 k_2 k_1
  + 4480 k_1 + 288 k_2^4 + 1952 k_2^3 + 4672 k_2^2 + 4480 k_2 + 1280)
  \d^2\varphi_1\d^2\varphi_1
  \\{}  
  + \tfrac{i\sqrt{k_1(k_2 + 2)}}{2\sqrt{2}}     
  (147 k_2 k_1^4 + 132 k_1^4 + 294 k_2^2 k_1^3 + 
  1292 k_2 k_1^3 + 880 k_1^3 + 147 k_2^3 k_1^2 + 
  1658 k_2^2 k_1^2 + 3720 k_2 k_1^2 + 2112 k_1^2\\{}
  + 498 k_2^3 k_1 + 2896 k_2^2 k_1 + 4320 k_2 k_1 + 2112 k_1
  + 408 k_2^3 + 1632 k_2^2 + 1744 k_2 + 704)  
  \d^2\varphi_2(\d\varphi_2)^2
  \\{}-
  \tfrac{\sqrt{k_1 k_2}}{4\sqrt{2}}  
  (57 k_2 k_1^4 + 60 k_1^4 + 114 k_2^2 k_1^3 + 428 k_2 k_1^3 + 
  304 k_1^3 + 57 k_2^3 k_1^2 + 836 k_2^2 k_1^2 + 
  1324 k_2 k_1^2 + 416 k_1^2\\{}
  + 468 k_2^3 k_1 + 2396 k_2^2 k_1+ 2256 k_2 k_1 - 64 k_1 + 528 k_2^3
  + 2000 k_2^2 + 1488 k_2 - 320)
  \d^2\varphi_2\d^2\varphi_2  
  \\{}-
  \tfrac{\sqrt{k_1 k_2}}{\sqrt{2}} 
      (k_1 + 2)(k_1 + k_2 + 5)
  (3 k_1 + 3 k_2 + 8)  
  (5 k_2 k_1 + 4 k_1 + 4 k_2 + 8)     
  \d^3\varphi_1\d\varphi_1
  \\{}+ 
  \tfrac{i\sqrt{k_1}}{\sqrt{2(k_2 + 2)}} 
  (k_2 + 1)  
  (2 k_1 + k_2 + 4)
  P
  \d^3\varphi_1\d\varphi_2
  \\{}+ 
  \tfrac{\sqrt{k_1 k_2}}{12\sqrt{2}(k_2 + 2)}    
  (15 k_2^2 k_1^4 + 42 k_2 k_1^4 + 24 k_1^4 + 
  30 k_2^3 k_1^3 + 130 k_2^2 k_1^3 + 296 k_2 k_1^3 + 
  192 k_1^3 + 15 k_2^4 k_1^2\\{}
  + 130 k_2^3 k_1^2 + 432 k_2^2 k_1^2 + 1044 k_2 k_1^2 + 800 k_1^2
  + 42 k_2^4 k_1 + 296 k_2^3 k_1 + 1044 k_2^2 k_1\\{}
  + 2192 k_2 k_1 + 1664 k_1 + 24 k_2^4 + 192 k_2^3
  + 800 k_2^2 + 1664 k_2 + 1280)    
  \d^4\varphi_1
  \\{}-
  \tfrac{i\sqrt{k_1}\,k_2}{12\sqrt{2(k_2 + 2)}}    
  (15 k_2 k_1^4 + 12 k_1^4 + 30 k_2^2 k_1^3 + 70 k_2 k_1^3 + 
  8 k_1^3 + 15 k_2^3 k_1^2 + 100 k_2^2 k_1^2 + 
  10 k_2 k_1^2 - 128 k_1^2\\{}
  + 42 k_2^3 k_1 + 138 k_2^2 k_1 - 
  88 k_2 k_1 - 224 k_1 + 24 k_2^3 + 56 k_2^2 - 16 k_2 - 64)    
  \d^4\varphi_2,
\end{multline}
where $K=k_1 + k_2 + 2$,
\begin{equation}
  P=37 k_2 k_1^2 + 44 k_1^2 + 37 k_2^2 k_1 + 192 k_2 k_1 + 
  176 k_1 + 44 k_2^2 + 176 k_2 + 176,
\end{equation}
and
\begin{multline}
  \xi^{-1} = 2\sqrt{\tfrac{5 k_1 k_2}{k_2 + 2}}
  \sqrt{(k_1 - 1)(k_2 - 1)}\sqrt{k_1 + 2}
  \sqrt{3 k_1 + 4} \sqrt{3 k_2 + 4}\times{}\\
  \times
  \sqrt{k_1 + k_2 + 2}\sqrt{k_1 + k_2 + 5}
  \sqrt{3 k_1 + 3 k_2 + 8} 
  \sqrt{P}.
\end{multline}

The next operator product expansion in the $W$ algebra is given by
\begin{multline}\label{W-appear}
  F(z)\,F(w) = \frac{c/4}{(z-w)^8} +
  \frac{2 T(w)}{(z-w)^6} +
  \frac{\d T(w)}{(z-w)^5} +
  \frac{a TT(w) + b \d^2T(w) + \gamma F(w)}{(z-w)^4} \\{}
  + \frac{a T \d T(w) + \beta \d^3T(w) +
    \frac{\gamma}{2}\d F(w)}{(z-w)^3} 
  + \frac{G(w) + \dots}{(z-w)^2} +\dots
\end{multline}
where $G$ is a new \textit{primary} dimension-6 operator (actually,
one of the two dimension-$6$ primary operators) that looks somewhat
more complicated than~\eqref{eq:F} (178 operator terms) and
\begin{gather*}
  a ={} \frac{42 (k_1+2) (k_2+2) (k_1+k_2+2)}{
    P},~
  b ={} \frac{3(k_2 k_1^2+8 k_1^2+k_2^2 k_1+12 k_2 k_1+32 k_1+8
    k_2^2+32 k_2+ 
    32)}{
    2P},\\
  \gamma ={} \frac{-3\sqrt{2/5}\,\cN}{ 
    \sqrt{(k_1-1)(k_2-1)(k_1+2)(k_2+2)(3 k_1+4)(3 k_2+4)
      K(k_1+k_2+5)(3 k_1+3 k_2+8)P}},\\
  \beta ={} \frac{2(11 k_2 k_1^2+25 k_1^2+11 k_2^2 k_1+69 k_2 k_1+100
    k_1+25 k_2^2 +100 k_2+100)}{3P},
\end{gather*}
\begin{multline*}
  \cN = 99 k_2^2 k_1^4-25 k_2 k_1^4
    -236 k_1^4+198 k_2^3 k_1^3+742 k_2^2 k_1^3-672 k_2 k_1^3-1888 k_1^3
    +99 k_2^4 k_1^2\\{}
    +742 k_2^3 k_1^2 + 576 k_2^2 k_1^2-4088 k_2 k_1^2
    -5168 k_1^2-25 k_2^4 k_1-672 k_2^3 k_1-4088 k_2^2 k_1 \\{}
    -8592 k_2 k_1-5568 k_1 -236 k_2^4-1888 k_2^3 -5168 k_2^2-
    5568 k_2-1792.
\end{multline*}

\section{The $\hD$ algebra} \label{app:D-all}
The Lie superalgebra $\D$, or $\mathfrak{osp}_\alpha(4|2)$, of the
dimension~$(9|8)$, can be described as follows.  Its bosonic
subalgebra is the direct sum of three $\SL2$ algebras generated by
$e^{(i)}$, $h^{(i)}$, and $f^{(i)}$, $i=1,2,3$, with the nonvanishing
commutators
\begin{equation}
  \begin{split}
    [e^{(i)}, f^{(j)}] ={}& 2 \delta_{i j} h^{(i)},\\
    [h^{(i)}, e^{(j)}] ={}&  \delta_{i j} e^{(i)},\\
    [h^{(i)}, f^{(j)}] ={}& -\delta_{i j} f^{(i)}.
  \end{split}
\end{equation}
The eight fermionic generators $\psi(\beta,\gamma,\delta)$, where
$\beta,\gamma,\delta=\p,\m$, are elements of the tensor product
$\oC^2\tensor\oC^2\tensor\oC^2$ of two-dimensional representations of
the three $\SL2$ algebras; the specific conventions are such that
\begin{equation}\label{1-psi}
  \begin{aligned}
    {}
    [e^{(1)},\psi(\m,\beta,\gamma)] ={}& -\psi(\p,\beta,\gamma),
    &\quad&
    [f^{(1)},\psi(\p,\beta,\gamma)] ={}& -\psi(\m,\beta,\gamma),\\
    [h^{(1)},\psi(\p,\beta,\gamma)] ={}& \thalf\psi(\p,\beta,\gamma),
    & &
    [h^{(1)},\psi(\m,\beta,\gamma)] ={}& -\thalf\psi(\m,\beta,\gamma)
  \end{aligned}
\end{equation}
and similarly for the other two $\SL2$ subalgebras (acting
respectively on the second and the third arguments of $\psi$).
Finally, to write the commutation relations for the fermions, we
introduce the ``spinor'' notation for $\SL2$:
\begin{equation}   
  O_{\p\p}=-2e,\qquad O_{\m\m}=2f,\qquad O_{\p\m}=O_{\m\p}=-2h.
\end{equation}
We then have
\begin{equation}
  [\psi(\beta_1,\beta_2,\beta_3),
  \psi(\gamma_1,\gamma_2,\gamma_3)]
  =
  \A_1 O_{\beta_1\gamma_1}^{(1)}
  \epsilon_{\beta_2\gamma_2}\epsilon_{\beta_3\gamma_3}
  + \A_2 O_{\beta_2\gamma_2}^{(2)}
  \epsilon_{\beta_1\gamma_1}\epsilon_{\beta_3\gamma_3}
  + \A_3 O_{\beta_3\gamma_3}^{(3)}
  \epsilon_{\beta_1\gamma_1}\epsilon_{\beta_2\gamma_2}
\end{equation}
where
\begin{equation}\label{A-Jacobi}
  \A_1 + \A_2 + \A_3 = 0.
\end{equation}
The algebra depends on $\A_1$, $\A_2$, and $\A_3$ modulo a nonzero
common factor.  Together with~\eqref{A-Jacobi} (which follows from the
Jacobi identities), this leaves \textit{one} independent parameter.
In the standard notation, we thus obtain the $\D$ algebra with
\begin{equation}\label{fix-alpha23}
  \alpha=-1-\frac{\A_2}{\A_3}.
\end{equation}
The Cartan matrix can be written as
\begin{equation}
  \begin{pmatrix}
    0 & 1 & -1-\alpha \\
    1 & 0 & \alpha \\
    -1-\alpha & \alpha & 0
  \end{pmatrix}.
\end{equation}
It is obviously a matter of convention which of the formulae
$\alpha=-1-\frac{\A_i}{\A_j}$ for any ordered pair $(\A_i,\A_j)$,
$i\neq j$, is used to define~$\alpha$ instead of~\eqref{fix-alpha23};
therefore, the discrete transformations
\begin{equation}\label{alpha-discrete}
  \alpha\mapsto-\alpha-1,\quad
  \alpha\mapsto\frac{1}{\alpha},\quad
  \alpha\mapsto-\frac{\alpha}{\alpha+1},\quad
  \alpha\mapsto-\frac{\alpha+1}{\alpha},\quad
  \alpha\mapsto-\frac{1}{\alpha+1}
\end{equation}
leave the $\D$ algebra invariant. 

The invariant form on $\D$ is given by (with $\epsilon_{\p\m}=1$)
\begin{equation}
  \begin{split}
    \langle e^{(i)}, f^{(j)}\rangle ={}& \frac{1}{\A_i}\,\delta_{i
      j},\\ 
    \langle h^{(i)}, h^{(j)}\rangle ={}&
    \frac{1}{2 \A_i}\,\delta_{i j},\\                 
    \langle\psi(\beta_1,\beta_2,\beta_3),
    \psi(\gamma_1,\gamma_2,\gamma_3)\rangle ={}&
    -2 \epsilon_{\beta_1\gamma_1}                     
    \epsilon_{\beta_2\gamma_2}
    \epsilon_{\beta_3\gamma_3}.
  \end{split}  
\end{equation}

We next construct the corresponding affine algebra: its generators are
given by $e^{(i)}_n$, $h^{(i)}_n$, $f^{(i)}_n$,
$\psi(\beta,\gamma,\delta)_n$, where $n\in\oZ$, and the central
element.  The invariant form depends on the overall scale of $\A_i$,
and therefore, the pair (algebra, invariant form) depends on
\textit{two} parameters.  In the corresponding affine algebra, one of
these parameters can be interpreted as the level.  Explicitly, the
affine-$\D$ commutation relations are given by
\begin{equation}
  \begin{split}
    [e^{(i)}_m, f^{(j)}_n] ={}& 2 \delta_{i j} h^{(i)}_{m+n} +
    \delta_{i j}\, \,m\,\delta_{m+n,0}\,\tfrac{1}{\A_i},\\
    [h^{(i)}_m, e^{(j)}_n] ={}&  \delta_{i j} e^{(j)}_{m+n},\\
    [h^{(i)}_m, f^{(j)}_n] ={}& -\delta_{i j} f^{(j)}_{m+n},\\
    [h^{(i)}_m, h^{(j)}_n] ={}& \delta_{ij}\,m\,
    \delta_{m+n,0}\,\tfrac{1}{2 \A_i},
  \end{split}
\end{equation}
by a straightforward ``affinisation'' of~\eqref{1-psi} and similar
formulae for the other two $\hSL2$ subalgebras, and by
\begin{multline}
  [\psi(\beta_1,\beta_2,\beta_3)_m,
  \psi(\gamma_1,\gamma_2,\gamma_3)_n]={}\\
  {}=
  \A_1 (O_{\beta_1\beta_1}^{(1)})_{m+n}
  \epsilon_{\beta_2\gamma_2}\epsilon_{\beta_3\gamma_3}
  + \A_2 (O_{\beta_2\gamma_2}^{(2)})_{m+n}
  \epsilon_{\beta_1\gamma_1}\epsilon_{\beta_3\gamma_3}
  + \A_3 (O_{\beta_3\gamma_3}^{(3)})_{m+n}
  \epsilon_{\beta_1\gamma_1}\epsilon_{\beta_2\gamma_2}\\*
  {} - 2 m\,\delta_{m+n,0}\,\epsilon_{\beta_1\gamma_1}
  \epsilon_{\beta_2\gamma_2}
  \epsilon_{\beta_3\gamma_3}.
\end{multline}

The levels $\varkappa_i=\frac{1}{\A_i}$ of the three $\hSL2$
subalgebras are therefore such that
\begin{equation}
  \frac{1}{\varkappa_1} + \frac{1}{\varkappa_2} +
  \frac{1}{\varkappa_3} = 0.
\end{equation}
Any of these can be called, conventionally, the level of $\hD$, for
example, $\varkappa=\varkappa_3=\frac{1}{\A_3}$.

The Sugawara energy-momentum tensor $\cT$ is given by
\begin{multline}\label{D-sug}
  \cT = 
  \A_1 e^{(1)}f^{(1)} + \A_2 e^{(2)}f^{(2)}+ \A_3 e^{(3)}f^{(3)} +
  \A_1h^{(1)}h^{(1)} + \A_2h^{(2)}h^{(2)} + \A_3h^{(3)}h^{(3)}\\
  {}+ 
  \thalf\psi(\p, \p,\p) \psi(\m,\m, \m) -
  \thalf\psi(\p, \p,\m) \psi(\m,\m, \p) \\
  {}- \thalf\psi(\p, \m,\p) \psi(\m,\p, \m) -
  \thalf \psi(\m,\p, \p)\psi(\p, \m,\m).  
\end{multline}
The central charge of the corresponding Virasoro algebra is equal
to~$1$.

\end{document}